\documentclass[preprint,12pt]{elsarticle}
\usepackage[a4paper, total={6in, 9in}]{geometry}

\usepackage{amsmath}
\usepackage{amssymb}
\usepackage{mathtools}
\usepackage{bm}
\usepackage{times}
\usepackage{colortbl}
\usepackage{tabularx}
\usepackage{float}
\usepackage{graphicx}
\usepackage[width=\linewidth, justification=centerlast]{caption}
\usepackage[justification=centerlast]{subcaption}
\usepackage[dvipsnames]{xcolor}
\usepackage{tikz}
\usepackage{color}
\usepackage{pgfplots}
\usepackage{cases}
\usepackage{bibentry}
\usepackage{multicol,multirow}
\usepackage{soul}
\soulregister\cite7
\soulregister\citep7
\soulregister\ref7
\usepackage{newtxtext}
\usepackage{newtxmath}
\usepackage{natbib}
\usepackage{hyperref}
\hypersetup{
  colorlinks = true,
  urlcolor = blue,
  citecolor= black,
}
\usepackage[displaymath]{lineno}
\usepackage{hhline}

\tikzset{
  basetraits/.style ={
    baseline=-0.6ex,
    very thick
  }
}

\tikzset{
  pics/solid/.style ={
    code = {
      \useasboundingbox (0,0) -- (0.54,0) ;
      \draw [pic actions] (0,0) -- (0.54, 0) ;
    }
  }
}

\DeclareRobustCommand{\soliddef}[1]{\tikz[basetraits] \draw[color=#1] pic{solid} ;}
\DeclareRobustCommand{\solid}{\soliddef{blue}}

\colorlet{cyan}[rgb]{cyan}
\DeclareRobustCommand{\dotteddef}[1]{\tikz[basetraits] \draw[color=#1, dotted] pic{solid} ;}
\DeclareRobustCommand{\dotted}{\dotteddef{cyan}}

\DeclareRobustCommand{\dasheddef}[1]{\tikz[basetraits] \draw[color=#1, dashed] pic{solid} ;}
\DeclareRobustCommand{\dashed}{\dasheddef{orange}}

\DeclareRobustCommand{\dashdotteddef}[1]{\tikz[basetraits] \draw[color=#1, dash dot] pic{solid} ;}
\DeclareRobustCommand{\chain}{\dashdotteddef{gray}}

\tikzset{
  pics/pic_triangle/.style ={
    code = {%
      \draw[color=#1] pic{solid} node[scale=1.2, rotate=-90] at (0.27,0) {\pgfuseplotmark{triangle*}} ;
    }
  }
}

\DeclareRobustCommand{\triangle}{%
  \tikz[basetraits] \draw (0,0) pic{pic_triangle={pink}} ;
}

\definecolor{EmeraldGreen}{RGB}{0, 128, 0}
\DeclareRobustCommand\square{%
   $\tikz[baseline=-0.6ex] \node[rectangle,draw,color=EmeraldGreen, very thick] (r) at (0,0) {};$%
}

\newcommand{\tb}{\mathbf{t}}

\newcommand{\x}{\mathbf{x}}
\newcommand{\w}{\boldsymbol{\theta}}
\newcommand{\PHI}{\boldsymbol{\Phi}}

\newcommand{\T}{\mathbf{T}}
\newcommand{\M}{\mathbf{M}}

\newcommand{\bb}{\mathbf{b}}

\newcommand{\DPS}{\displaystyle}
\newcommand{\derp}[2]{\frac{\partial #1}{\partial #2}}

\DeclareMathOperator*{\argmax}{arg\,max}

\newcommand{\ie}{\emph{i.e.}}

\newcommand{\expcomp}{\eta} 

\newcommand{\St}{\mathbf{S}} 
\newcommand{\Ot}{\boldsymbol{\Omega}} 
\newcommand{\vel}{U} 
\newcommand{\pres}{P} 

\newcommand{\muline}[1]{\begin{tabular}[c]{@{}c@{}} #1 \end{tabular}}

\newcommand{\Xavier}[1]{{\color{blue} #1}}

\journal{...}
\RequirePackage[normalem]{ulem} 
\RequirePackage{color}\definecolor{RED}{rgb}{1,0,0}\definecolor{BLUE}{rgb}{0,0,1} 
\providecommand{\DIFaddtex}[1]{{\protect\color{blue}\uwave{#1}}} 
\providecommand{\DIFaddbegin}{} 
\providecommand{\DIFaddend}{} 
\providecommand{\DIFdelbegin}{} 
\providecommand{\DIFdelend}{} 
\providecommand{\DIFaddbeginFL}{} 
\providecommand{\DIFaddendFL}{} 
\providecommand{\DIFdelbeginFL}{} 
\providecommand{\DIFdelendFL}{} 
\providecommand{\DIFadd}[1]{\texorpdfstring{\DIFaddtex{#1}}{#1}} 
\newcommand{\DIFscaledelfig}{0.5}
\RequirePackage{settobox} 
\RequirePackage{letltxmacro} 
\newsavebox{\DIFdelgraphicsbox} 
\newlength{\DIFdelgraphicswidth} 
\newlength{\DIFdelgraphicsheight} 
\LetLtxMacro{\DIFOincludegraphics}{\includegraphics} 
\newcommand{\DIFaddincludegraphics}[2][]{{\color{blue}\fbox{\DIFOincludegraphics[#1]{#2}}}} 
\newcommand{\DIFdelincludegraphics}[2][]{
\sbox{\DIFdelgraphicsbox}{\DIFOincludegraphics[#1]{#2}}
\settoboxwidth{\DIFdelgraphicswidth}{\DIFdelgraphicsbox} 
\settoboxtotalheight{\DIFdelgraphicsheight}{\DIFdelgraphicsbox} 
\scalebox{\DIFscaledelfig}{
\parbox[b]{\DIFdelgraphicswidth}{\usebox{\DIFdelgraphicsbox}\\[-\baselineskip] \rule{\DIFdelgraphicswidth}{0em}}\llap{\resizebox{\DIFdelgraphicswidth}{\DIFdelgraphicsheight}{
\setlength{\unitlength}{\DIFdelgraphicswidth}
\begin{picture}(1,1)
\thicklines\linethickness{2pt} 
{\color[rgb]{1,0,0}\put(0,0){\framebox(1,1){}}}
{\color[rgb]{1,0,0}\put(0,0){\line( 1,1){1}}}
{\color[rgb]{1,0,0}\put(0,1){\line(1,-1){1}}}
\end{picture}
}\hspace*{3pt}}} 
} 
\LetLtxMacro{\DIFOaddbegin}{\DIFaddbegin} 
\LetLtxMacro{\DIFOaddend}{\DIFaddend} 
\LetLtxMacro{\DIFOdelbegin}{\DIFdelbegin} 
\LetLtxMacro{\DIFOdelend}{\DIFdelend} 
\DeclareRobustCommand{\DIFaddbegin}{\DIFOaddbegin \let\includegraphics\DIFaddincludegraphics} 
\DeclareRobustCommand{\DIFaddend}{\DIFOaddend \let\includegraphics\DIFOincludegraphics} 
\DeclareRobustCommand{\DIFdelbegin}{\DIFOdelbegin \let\includegraphics\DIFdelincludegraphics} 
\DeclareRobustCommand{\DIFdelend}{\DIFOaddend \let\includegraphics\DIFOincludegraphics} 
\LetLtxMacro{\DIFOaddbeginFL}{\DIFaddbeginFL} 
\LetLtxMacro{\DIFOaddendFL}{\DIFaddendFL} 
\LetLtxMacro{\DIFOdelbeginFL}{\DIFdelbeginFL} 
\LetLtxMacro{\DIFOdelendFL}{\DIFdelendFL} 
\DeclareRobustCommand{\DIFaddbeginFL}{\DIFOaddbeginFL \let\includegraphics\DIFaddincludegraphics} 
\DeclareRobustCommand{\DIFaddendFL}{\DIFOaddendFL \let\includegraphics\DIFOincludegraphics} 
\DeclareRobustCommand{\DIFdelbeginFL}{\DIFOdelbeginFL \let\includegraphics\DIFdelincludegraphics} 
\DeclareRobustCommand{\DIFdelendFL}{\DIFOaddendFL \let\includegraphics\DIFOincludegraphics} 
\RequirePackage{listings} 
\RequirePackage{color} 
\lstdefinelanguage{DIFcode}{ 
  moredelim=[il][\color{red}\sout]{\%DIF\ <\ }, 
  moredelim=[il][\color{blue}\uwave]{\%DIF\ >\ } 
} 
\lstdefinestyle{DIFverbatimstyle}{ 
	language=DIFcode, 
	basicstyle=\ttfamily, 
	columns=fullflexible, 
	keepspaces=true 
} 
\lstnewenvironment{DIFverbatim}{\lstset{style=DIFverbatimstyle}}{} 
\lstnewenvironment{DIFverbatim*}{\lstset{style=DIFverbatimstyle,showspaces=true}}{} 

\begin{document}
\nolinenumbers
\begin{frontmatter}
  \title{Space-dependent Aggregation of Stochastic Data-driven Turbulence Models}
  \author[inst1]{S. Cherroud}
  \affiliation[inst1]{organization={DynFluid laboratory, Arts \& Métiers ParisTech},
    addressline={151 boulevard de l'Hôpital}, 
    city={Paris},
    postcode={75013}, 
    country={France}}
  \author[inst1]{X. Merle}

  \author[inst2]{P. Cinnella}
  \author[inst1]{X. Gloerfelt}
  \affiliation[inst2]{organization={Institut Jean Le rond d'Alembert, Sorbonne Université},
    addressline={4 Place Jussieu}, 
    city={Paris},
    postcode={75252},
    country={France}}

  \begin{abstract}
      A stochastic Machine-Learning approach is developed for data-driven Reynolds-Averaged Navier-Stokes (RANS) predictions of turbulent flows, with quantified model uncertainty. This is done by combining a Bayesian symbolic identification methodology for learning stochastic RANS model corrections for selected classes of flows (expert models), and a Mixture-of-Experts methodology that aggregates their predictions through space-dependent weighting functions that depend on a set of local flow features. The expert models are learned using the recently proposed SBL-SpaRTA algorithm \cite{cherroud2022sparse}, which generates sparse analytical expressions of the corrective terms with model parameters described by probability distributions. The learned models are naturally interpretable and equipped with an intrinsic measure of uncertainty. They outperform the baseline RANS model for flow classes similar to those used for their training, but their generalization to radically different flows is not warranted.  With the aim of quantifying the predictive uncertainty associated with the data-driven models while improving predictive accuracy and generalization capabilities, a space-dependent model aggregation technique (XMA) is then adopted \cite{MZB2023}. A gating function, which assigns each model a performance score (weight) based on a vector of local flow features, is trained alongside the expert models. The weights can be interpreted as the probability that a candidate model will outperform its competitors given the flow behavior at a given location. Predictions of unseen flows are then formulated as a locally weighted average of the stochastic solutions of the competing expert models. Furthermore, a prediction uncertainty estimate is obtained by propagating through the RANS solver the models' posterior parameter distributions and by evaluating the inter-model prediction variance. The expectancy of the XMA prediction is found to be significantly more accurate than the baseline deterministic solution and the individual solutions of the experts for a set of well-documented benchmark flows not included in the training set, while providing consistent estimates of the predictive variance.
  \end{abstract}



\end{frontmatter}


\section{Introduction}
Turbulence models are a crucial component of Computational Fluid Dynamics (CFD) solvers, widely employed for flow analysis and design in fluids engineering. CFD of real-world problems is still largely based on Reynolds-Averaged Navier-Stokes (RANS) equations supplemented with a suitable closure model for the so-called Reynolds stresses, representative of the influence of turbulent scales on the mean flow. Many different models have been developed over the years, with varying degrees of success in predicting extended ranges of flows. The reader is referred, e.g., to \cite{spalart2000,spalart2015,durbin2018some} and to the books of \cite{pope2000turbulent,wilcox2006turbulence} for an overview of RANS turbulence models. RANS turbulence modeling continues to be the subject of ongoing research, as exemplified by a recent international symposium \cite{rumsey2022nasa}. Despite continuous research efforts though several decades,  "no class of models has emerged as clearly superior, or clearly hopeless" until now \cite{spalart2000} . Clearly, RANS models suffer from several shortcomings for complex flow configurations involving turbulence nonequilibrium, strong gradients, separations, shocks, 3D effects, etc. An important source of error lies in the constitutive assumption known as the linear eddy viscosity or Boussinesq hypothesis, which is used in the vast majority of models for industrial use. So-called linear eddy viscosity models (LEVM) postulate a linear relationship between the Reynolds stress tensor and the mean strain rate tensor, a property that is not fully verified even for relatively simple flows \cite{wilcox2006turbulence,schmitt2007boussinesq}. A large part of the turbulence modeling literature during the last three decades reports attempts to upgrade the baseline LEVM by adding nonlinear terms suited to sensitize the model to curvature effects or to improve its anisotropy.
Examples are given by the SARC (Spalart--Allmaras with Rotation and Curvature, \cite{spalart_shur1997}), non-linear models \citep{speziale1987}, elliptic relaxation models \citep{durbin1991}, algebraic Reynolds Stress models \citep{rodi1976algebraic} or explicit algebraic Reynolds Stress models (\cite{pope1975more,gatski1993explicit,wallin2000explicit}), and full Reynolds Stress Models \citep{speziale1995}. The latter require the solution of additional transport equations for the Reynolds stress components plus an equation for a quantity allowing to determine a turbulent scale. 
Unfortunately, the balance accuracy / robustness / computational cost of such more complex models has prevented a widespread use in CFD applications. In addition, more complex models typically involve a larger number of adjustable closure coefficients, typically calibrated against experimental or numerical data for so-called "canonical flows", \emph{i.e.}, simple turbulent flows representative of some elementary turbulent dynamics. However, 1) it is not always possible to determine closure coefficients that are simultaneously optimal for all canonical flows (as exemplified by, e.g., the so-called round/plane jet anomaly \cite{wilcox2006turbulence}); 2) the calibration data are affected by observational uncertainties that propagate to the closure coefficients; and 3) the final values retained in some models are not even the best fit to the data, but rather a compromise with respect to other requirements, e.g. numerical robustness. A discussion of uncertainties associated with turbulence models can be found in \cite{xiao2019quantification}.
\\

An alternative line of research has consisted in the development of so-called scale-resolving approaches (Large Eddy Simulation (LES), wall-modelled LES, and hybrid RANS / LES), which directly resolve a more or less extended range of turbulent structures while modeling scales smaller than a certain filter length. Such approaches have been successful for increasingly realistic flow configurations. However, their computational cost remains significant and even prohibitive for applications requiring multiple simulations, such as massive parametric studies, optimization or Uncertainty Quantification (UQ).
\\

In recent years, there has been a growing interest in applying Machine-Learning (ML) techniques for turbulence modeling. ML can be used to analyze large amounts of data and discover non-trivial patterns. Since all RANS models involve some degree of empiricism \citep{spalart2000,spalart2015}, including those that were initially derived from exact manipulations of the Navier--Stokes equations, the use of ML has emerged during the last decade as a natural way to systematize the development of RANS models and discover improved RANS formulations for more complex flows (see the reviews of \cite{duraisamy2019turbulence,xiao2019quantification,duraisamy2020perspectives,sandberg2022review}).

Early studies of so-called data-driven RANS models focused on quantifying and reducing model uncertainties using interval analysis or statistical inference tools. This was achieved either by directly perturbing the Reynolds stress anisotropy tensor computed with a basic LEVM model \cite{emory2013modeling,gorle2013framework,thompson2019eigenvector} or by treating the model closure coefficients as random variables endowed with probability distributions \cite{platteeuw2008uncertainty,cheung2011bayesian,edeling2014bayesian,margheri2014epistemic}.
The first approach accounts for structural uncertainties in the constitutive relation of the Reynolds stresses, while the second considers only parametric uncertainties associated with the closure coefficients. 

With the aim of reducing modeling inadequacies, data-driven  methods for turbulence modeling have  been introduced in recent years, mostly relying on supervised ML. Examples of early contributions can be found in \cite{tracey2013application,parish2016paradigm}, who proposed field inversion to learn corrective terms for the turbulent transport equations, along with ML to express the correction as a black-box function of selected flow features and to extrapolate it to new flows. Other contributions to the field-inversion and ML approach can be found, e.g., in \cite{ferrero2020field,volpiani2021machinePRF,volpiani2022neuralIJHFF}. One of the advantages of field inversion is that the (goal-oriented) correction can be inferred from sparse data or even global performance parameters. 
On the other hand, the efficacy of the learned correction is strongly dependent on the features used to map it to unseen flow cases \citep{srivastava2021generalizable}. Even when an appropriate set of feaures is available, the learned corrections tend to lack generality, and cannot be applied to flows significantly different from those used to learn the correction \citep{rumsey2022search}.
The seminal work of \cite{ling2015evaluation} introduced a novel neural network architecture (Tensor Basis Neural Network, TBNN) that allows frame invariance constraints to be incorporated into the learned Reynolds stress anisotropy correction. The idea is to project the correction term onto a minimal integrity basis, as in the extended eddy viscosity model of \cite{pope1975more}, leading to a form of generalized Explicit Algebraic Reynolds Stress model, whose function coefficients are regressed from high-fidelity data using ML. 
Pope's invariant representation has also been employed recent contributions are given by \cite{kaandorp2020data} who used tensor-basis Random Forests to learn data-driven corrections of the Reynolds stress tensor, and \cite{jiang2021interpretable} who proposed a general principled framework for deriving deep learning turbulence model corrections using deep neural network (DNN) while embedding physical constraints and symmetries. 
Refs \cite{xiao2016quantifying,wu2016bayesian,wu2017priori,wu2018physics,wu2019physics} combined ML classification techniques for identifying regions of high RANS modeling uncertainty with ML regression techniques for inferring model corrections from data and  predicting new configurations. The procedure, initially relying on the assimilation of full high-fidelity fields, has been subsequently extended to the assimilation of sparse data by using end-to-end differentiation \cite{strofer2021end}. In \cite{zhou2022frame}, the use of vector cloud ML upholds the desired invariance properties of constitutive models, accurately reflects the physical region of influence, and can be applied to various spatial resolutions; however, it still needs to integrate the informations about the turbulent length- and velocity-scales via transport equations to provide a better description of the Reynolds stresses.

The aforementioned approaches to modeling turbulence from data use so-called black-box ML, such as neural networks or Random Forests. They allow a flexible approximation of complex functional relationships, but do not provide an explicit, physically interpretable mathematical expression for the learned correction. Recent attempts to interpret ML-augmented turbulence models rely on nonlinear sensitivity analysis tools, such as Shapley factor analyses \citep{he2022explainability}. 
An interesting alternative is represented by so-called open-box ML approaches, which consist in selecting explicit mathematical expressions and/or operators from a large pre-defined dictionary to build a suitable regressor for the data. 
Examples of open-box ML include Genetic Programming (GEP) \citep{weatheritt2016novel} and symbolic identification \citep{schmelzer2020discovery,beetham2020formulating,mandler2022realizable}. 
Although less expressive than black-box ML, due to the rapidly escalating complexity of the search procedure in large mathematical operator dictionaries, the open-box models have the merit of providing tangible mathematical expressions that can be easily integrated into existing CFD solvers and interpreted in the light of physical considerations. Both a priori and "CFD-in-the-loop" training methods have been proposed \citep{zhao2020rans,saidi2022}, the latter allowing the use of incomplete data, at the cost of solving a large optimization problem.

Regardless of their formulation and training procedure, both black-box and open-box suffer from common drawbacks. First, the learned corrections are generally non-local, which means that they can alter the predictions of RANS models even where the baseline LEVM already gives good results. Second, such models tend to perform well only for narrow classes of flows and operating conditions, which means that they often need to be retrained whenever a new flow configuration needs to be addressed. Third, most data-driven model do not provide any information about the prediction uncertainty, which can be large when the model is applied to unseen datasets.

An attractive approach for improving generalization to a wider set of flow while quantifying the predictive uncertainty consists in using multi-model ensemble techniques. One of the first applications of such methodology to turbulence modeling can be found in \cite{edeling2014predictive}, where the BMSA methodology was used to combine the solutions of a set of competing  LEVM models calibrated to different data sets (scenarios). In BMSA, each component model is used to make a prediction of a new flow, and an aggregated estimate of the solution is obtained as a linear combination of the competing model predictions weighted by the posterior model probabilities. The solution variance can also be evaluated as a function of the uncertain model parameters, model structure, and choice of the calibration data, providing an estimate of the prediction confidence intervals. BMSA has been successfully applied to  provide stochastic predictions for a variety of flows,    including     3D    wings     \citep{edeling2018bayesian}    and     compressor    cascades \citep{MZB2020,MZB2022}.
BMSA, and Bayesian Model  Averaging (BMA) \citep{draper1995assessment,hoeting1999bayesian} from which
it originates, may be interpreted as stochastic  variants of a multi-model framework called Model Aggregation
\citep{stoltz2010agregation,  devaine2013forecasting,deswarte2019sequential}. Such methods  combine multiple predictions stemming  from various models --also termed experts or forecasters--  to provide a global, enhanced solution. The above-mentioned methods, however, assign to each model the same weight throughout the domain. Since, in practice, models perform better or worse depending on the local flow physics, a better strategy consists in assigning higher weights to the best performing models in each region. 
Other   classes  of   ensemble  methods   allow  space-varying   weights. Specifically, so-called Mixture-of-Experts models \citep{yuksel_twenty_2012}  or  Mixture Models, softly  split the  input feature space  (covariate space) into partitions where  the locally best-performing models are assigned  higher weights.  The soft partitioning is  accomplished through parametric gating functions, or a  network of hierarchical gating functions \citep{JordanJacobs_1994expertregions}, that rank the  model outputs with probabilities.

In this spirit, some of the present authors recently proposed in Ref. \cite{MZB2023} (to which we refer for a more complete literature review on ensemble models) a method for spatially combining the predictions of a set of turbulence models, called space-dependent Model Aggregation (XMA). For that purpose, a cost function is introduced to evaluate the local model performance with respect to some training data, which is used to build the model weights. To make predictions, the weights are regressed in a space of flow features (representative of different flow phenomena) by using Random Forests. Similar to BMSA, an estimate of the predictive uncertainty can be inferred by measuring the degree of agreement of the component model predictions. In \cite{MZB2023} the XMA methodology was successfully applied to aggregate a set of well-known LEVM from the literature to predict flows through a compressor cascade at various operating conditions. 

In the present work we build on the XMA methodology of \cite{MZB2023}, but instead of combining a set of  "on-the-shelf" LEVMs, we first train a set of data-driven models using open-box ML. Specifically, the Symbolic Bayesian Learning Sparse Regression of Turbulent stress Anisotropy (SBL-SpaRTA) of \cite{cherroud2022sparse} is used to learn customized model corrections for various flow datasets, including fully developed channel flows, a near-sonic axisymmetric jet, a turbulent boundary layer subject to different adverse pressure gradients, and separated flow cases. SBL-SpaRTA combines the efficient Sparse Bayesian Learning (SBL) algorithm of \cite{tipping2001sparse} and the SpaRTA (Sparse Regression of Turbulent stress Anisotropy) symbolic-regression approach of \cite{schmelzer2020discovery} to learn analytical model corrections whose parameters are described by probability distributions, i.e. are endowed with an intrinsic measure of uncertainty. The stochastic customized models, or "experts", are then aggregated by training weighting functions that depend on a set of flow features, representative of typical physical mechanisms encountered in incompressible turbulent flows. Finally, the XMA model is used to predict the solution expectancy and variance for flow cases not used for training. 

The structure of the paper is as follows. In Section \ref{SBLTMC}, we briefly recall the SBL-SpaRTA methodology. 
The XMA model aggregation approach is presented in Section \ref{sec:spacedependent}. In section \ref{modeltraining} is presented the numerical training of the SBL-SpaRTA and of the XMA weights. Finally, in Section \ref{sec:resultsxma},  we illustrate the performance of XMA, first for a set of training flow cases and then for a set of well-documented unseen turbulent flow configurations.

\section{Methodology}
We consider the steady incompressible RANS equations:
\begin{equation}
  \begin{dcases}\label{RANS}
    \frac{\partial U_i}{\partial x_i} & = 0 \\
    U_j\frac{\partial U_i}{\partial x_j} & = -\frac{1}{\rho}\frac{\partial P}{\partial x_i} + \frac{\partial }{\partial x_j}\left( \nu\frac{\partial U_j}{\partial x_J}-\tau_{ij}\right)%
  \end{dcases}
\end{equation}
with $U_i$ the $i$-th mean velocity component, $P$ the mean pressure, $\rho$ the fluid density and $\nu$ the kinematic viscosity. 
In the following, we seek closure models for the so-called Reynolds stress tensor $\tau_{ij}=\langle u'_i u'_j\rangle$, with $u'_i$ the $i$-th fluctuating velocity component. The most widely used turbulence models, LEVMs, are based on the so-called 'Boussinesq hypothesis', which postulates a linear dependence between the Reynolds stress tensor and the mean strain rate tensor $S_{ij}$:
\begin{equation}\label{Bouss}
  \tau_{ij} = 2k \left(\frac{1}{3}\delta_{ij} - \frac{\nu_t}{k}S_{ij}\right)
  ,\quad S_{ij} = \frac{1}{2} \left(\derp{U_i}{x_j} + \derp{U_j}{x_i}\right)
\end{equation}
with $k$ the turbulent kinetic energy (TKE) and $\nu_t$ the eddy viscosity coefficient. This assumption leads to shift the modeling target to $\nu_t$, which is often computed by solving auxiliary transport equations for well-chosen turbulent properties, such as in the $k$-$\omega$ SST model \citep{menter1992improved} used in the remainder of this study.\\

In practice, the constitutive relation (\ref{Bouss}) is inaccurate for flows with rapid distorsion, large pressure gradients, separations, or for genuinely 3D flows, and needs to be improved. Following \cite{schmelzer2020discovery},
we correct $\tau_{ij}$ by introducing a second-order symmetric and traceless tensor $b_{ij}^{\Delta}$,\,called extra-anisotropy, such that:
\begin{equation}\label{constitutive}
  \tau_{ij} = 2k \Biggl( \frac{1}{3} \delta_{ij} \underbrace{- \frac{\nu_t}{k}S_{ij} + b_{ij}^{\Delta}}_{b_{ij}}\Biggr)
\end{equation}
Our goals is then to learn the extra-anisotropy $b_{ij}^{\Delta}$ from high-fidelity data. For that purpose, full fields of high-fidelity Reynolds stresses $\tau_{ij}$ and turbulent kinetic energy  $k$ are used to obtain target data $b_{ij}^{\Delta,HF}$ from equation (\ref{constitutive}). In addition, high-fidelity data are also used to compute the baseline modelled anisotropy  $- \frac{\nu_{t,HF}}{k_{HF}}S_{ij,HF}$. While  $S_{ij,HF}$ is easily computed from the high-fidelity velocity components $U_{i,HF}$, several choices are possible for $\nu_{t,HF}$.
If the $k-\omega$ SST model \citep{menter1992improved} is used as the baseline, $\nu_t$ is evaluated from auxiliary transport equations $k$ and the specific dissipation rate $\omega$. 
While the latter is sometime available in high-fidelity databases (at least for DNS databases), using such "true" $\omega$ has been found to possibly lead to severe inconsistencies with the value returned by the modeled $\omega$ equation, which is heavily tuned to give a good estimate of the shear stress in a boundary layer \citep{weatheritt2017development}. To avoid this, we follow instead the $k$-corrective frozen-RANS procedure of \cite{schmelzer2020discovery}, which consists in passively solving the $k-\omega$ SST transport equations by using the high-fidelity mean flow quantities and turbulent kinetic energy $k$. To improve the consistency of the data-driven model with the exact kinetic energy budget, the modeled $k$ equation is  modified by injecting a residual term, denoted $R$, which accounts for the discrepancy between the exact and the modeled terms; a counterpart of the $R$ term is also added to the $\omega$ equation. 

In practice, the following equations are solved for $R$ and $\omega$ by computing all other terms from high fidelity data:
\begin{equation}
  \begin{dcases}
    U_{j} \derp{k}{x_{j}} = P_{k}+ R -\beta^{*} k\omega+ \derp{}{x_{j}} \left((\nu
    +\sigma_{k} \nu_{t}) \derp{k}{x_j} \right) \\ 
    U_{j}\derp{\omega}{x_j} =\frac{\gamma}{\nu_{t}}(P_{k} + R) -\beta\omega^{2}+
    \derp{}{x_j} \left((\nu +\sigma_{\omega} \nu_{t}) \derp{\omega}{x_j} \right) +2(1-F_1)\sigma_{\omega 2} \frac{1}{\omega}\frac{\partial k}{\partial x_i}\frac{\partial \omega}{\partial x_i}\\
    \nu_t=\DPS\frac{a_1 k}{\max\left(a_1\omega, \Omega F_1\right)}
  \end{dcases}
\end{equation}
where $a_1, \beta,\,\beta^*,\,\gamma, \,\sigma_k,\,\sigma_\omega,\sigma_{\omega 2}$ are closure coefficients, $F_1$ and $F_2$ are blending functions \cite{menter1992improved}, and $\Omega$ is the vorticity magnitude. Furthermore,
\begin{equation}\label{datacuration:production}
 P_k := -\tau_{ij}\derp{U_i}{x_j} = 2\nu_tS^2 - 2k b_{ij}^{\Delta}\derp{U_i}{x_j} %
\end{equation}
The frozen-RANS value for $\omega$ and $k$ is used to iteratively determine the "high-fidelity" eddy viscosity $\nu_{t,HF}$, hence $b_{ij}^{\Delta,HF}$, and a turbulent timescale is estimated as $1/\omega$.

Upon convergence of $k$-corrective frozen-RANS, pre-processing of high-fidelity data ultimately provides two target terms, $b_{ij}^{\Delta}$ and $R$, which are modelled using open-box machine learning.

The extra-anisotropy is assumed to be a function of the mean velocity gradient only, and projected onto a minimal integrity basis of ten tensors and five scalar invariants initially introduced in Ref. \cite{pope1975more}. For 3D problems, the adopted Reynolds stress representation writes:
\begin{equation}
  b^{\Delta}_{ij}=\sum_{l=1}^{10} \alpha_{l}^{\Delta}(I_{1},...,I_{5})  T_{ij}^{(l)}
 \end{equation}
For  2D flows as those considered in this work, only the first three tensors are linearly independent, and only two scalar invariants are nonzero, so that:
\begin{equation}\label{bdelta}
  b^{\Delta}_{ij}= \sum_{l=1}^{3} \alpha_{l}^{\Delta} (I_{1},I_{2}) T_{ij}^{(l)} 
 \end{equation}
where :
\begin{equation}
\begin{dcases}\label{tensors}
T_{ij}^{(1)}= S_{ij}^* \\
T_{ij}^{(2)}= S_{il}^*\Omega_{lj}^* - \Omega_{il}^* S_{lj}^* \\
T_{ij}^{(3)} = S_{il}^* S_{lj}^* - \frac{1}{3} \delta_{ij} S_{mn}^* S_{mn}^* \\
I_{1} = S_{mn}^* S_{mn}^* \\
I_{2} = \Omega_{mn}^* \Omega_{mn}^*
\end{dcases}
\quad\text{and}\quad
\begin{dcases}
\vspace{0.2cm}
S_{ij}^* = \frac{S_{ij}}{\omega} =\frac{1}{2\omega}\left(\derp{U_i}{x_j} + \derp{U_j}{x_i}\right) \\
\Omega_{ij}^* = \frac{\Omega_{ij}}{\omega} =\frac{1}{2\omega}\left(\derp{U_i}{x_j} - \derp{U_j}{x_i}\right)
\end{dcases},
\end{equation}
In the above, $S_{ij}^*$ is a non-dimensional strain rate tensor and $\Omega_{ij}^*$ a non-dimensional rotation rate tensor. The corrected constitutive relation given by Equations \eqref{constitutive} and \eqref{bdelta} has the general form of an Explicit Algebraic Stress model \citep{gatski1993explicit} with function coefficients to be learned from data. 

A modeling ansatz for $R$ is also needed. This is expressed as a "production-like" term, \emph{i.e.} as the inner product of a second-order symmetric tensor, noted $b_{ij}^R$ and the mean velocity gradient:
\begin{equation}\label{bR}
R= b^{R}_{ij}\frac{\partial U_i}{\partial x_j},\quad
  b^{R}_{ij}= \sum_{l=1}^{3} \alpha_{l}^{R} (I_{1},I_{2}) T_{ij}^{(l)}
\end{equation}
Finally, $\alpha_l^{\Delta}$ and $\alpha_l^{R}$ are sought as linear combinations of vectors of the set $\boldsymbol{\mathcal{B}}$, a library of monomials of the invariants $I_1$ and $I_2$:
\begin{equation}
\boldsymbol{\mathcal{B}} = \Bigl\{ I_1^{m},\ I_2^{n},\ I_1^{p}I_2^{q}\;\bigl\vert\; m,n \in \mathbb{N},\; p,q \in \mathbb{N}^{*}, \ 0 \leq m,n \leq 9, \ 2\leq p+q \leq 4 \Bigr\} 
\end{equation}
leading to 25 candidate terms for each function ($\alpha_l^{\Delta}$ and $\alpha_l^R$), \emph{i.e.} a total of $25\times 3 =75$ ($k=\{1,2,3\}$) candidate functions for each tensor ($b_{ij}^{\Delta}$ and $b_{ij}^{R}$).

Given a high-fidelity dataset with data available at $N$ observation points, the learning problem is then rewritten in the compact form \citep{cherroud2022sparse}:
\begin{equation}\label{learning:problem}
\begin{cases}
\DPS \tb^{\Delta} =\PHI_{b^{\Delta}}\w_{b^{\Delta}} \\
\DPS \tb^R = \PHI_{R} \w_{R} 
\end{cases}
\end{equation}
where
\begin{equation}
\begin{cases}
\tb^{\Delta}=2k(b^{\Delta}_{11|n=1},...,b^{\Delta}_{11|n=N},......,b^{\Delta}_{33|n=1},...,
b^{\Delta}_{33|n=N})^{T} 
\\
\tb^{R}=(R_{|n=1},...,R_{|n=N})^{T} \\
\end{cases}
\end{equation}
are two high-fidelity target value vectors of size $4N$ (in 2D) and $N$, respectively, and $\PHI_{b^{\Delta}}$ and $\PHI_{R}$ are two matrices of candidate function terms evaluated at the $N$ observation points:
\begin{equation}
\begin{dcases}
\PHI_{b^{\Delta}} = 2k
\begin{bmatrix}
T_{11|n=1}^{(1)} & I_1 T_{11|n=1}^{(1)} &... & I_1^{2} I_{2}^{2}T_{11|n=1}^{(3)} \\
T_{11|n=2}^{(1)} & I_1 T_{11|n=2}^{(1)} & ... & I_1^{2} I_{2}^{2}T_{11|n=2}^{(3)} \\
... & ... & ... & ... \\
T_{33|n=N}^{(1)} & I_1 T_{33|n=N}^{(1)} & ... & I_1^{2} I_{2}^{2}T_{33|n=N}^{(3)}
\end{bmatrix} \\
\PHI_{R} = 2k
\begin{bmatrix} 
\left\{T_{ij}^{(1)} \partial_{j} U_{i}\right\}_{n=1} & I_1 \left\{T_{ij}^{(1)}\partial_{j} U_{i}\right\}_{n=1} & ... &
I_1^{2}I_{2}^{2} \left\{T_{ij}^{(3)} \partial_{j} U_{i}\right\}_{n=1} \\
... & ... & ... & ... \\
\left\{T_{ij}^{(1)} \partial_{j} U_{i}\right\}_{n=N} & I_1 \left\{T_{ij}^{(1)}\partial_{j} U_{i}\right\}_{n=N} & ... &
I_1^{2} I_{2}^{2} \left\{T_{ij}^{(3)} \partial_{j} U_{i}\right\}_{n=N} 
\end{bmatrix}
\end{dcases}
\end{equation}
$\PHI_{b^{\Delta}}$ and $\PHI_{R}$ are of size $4N\times 75$ and $N\times 75$, respectively. Finally, $\w_{b^{\Delta}}$ and $\w_R$ are the two unknown vectors of decomposition coefficients of the $\alpha_k^{\Delta}$'s and the $\alpha_k^{R}$'s on $\boldsymbol{\mathcal{B}}$, and are of size $75$ each one.

The learning problem (\ref{learning:problem})
is solved by means of the Sparse Bayesian Learning (SBL) algorithm, initially proposed in \cite{tipping2001sparse}. SBL is a method for finding sparse stochastic solutions to generalized linear inverse problems. SBL is based on Bayesian inference and uses a structured set of hierarchical prior distributions to promote sparse solutions, \emph{i.e.} model parameter vectors with a small number of non-zero components.

Specifically, the SBL algorithm seeks a model parameter vector $\w=(\theta_1,...,\theta_M)^T$ for linking a set of input-output pairs $\{\x_{n},t_{n} \}_{n=1}^{N}$ via a probabilistic formulation that assumes that the targets $\mathbf{t} = (t_1,...,t_N)^T$ are sampled from a generalized linear model with additive noise $\boldsymbol{\epsilon}$:
  \begin{equation}
    t_n(\x_n;\w) = \sum_{i=1}^M \Phi_i(\x_n) \theta_i + \epsilon_n,\;n=1,\ldots N
\end{equation}
where the $\Phi_i$ are the candidate functions, $\PHI$ is the design matrix such that $\{\PHI\}_{ni} = \Phi_i(\x_n)$ and $\boldsymbol{\epsilon}= (\epsilon_1,...,\epsilon_N)^T$ is a vector of independent noise processes assumed to be Gaussian, with zero mean and variance $\sigma^2$. In our framework, $\PHI$ could be either $\PHI_{b^{\Delta}}$ or $\PHI_{R}$ and $\w$ could be $\w_{b^{\Delta}}$ or $\w_R$ respectively.
The parameters $\w$ are initially assigned large Gaussian distributions centered at 0, the variances of which are treated as hyperparameters to be learned alongside $\w$ from the data. Following \cite{cherroud2022sparse}, the hyperparameters are assigned in their turn a Laplace hyperprior, which enforces model sparsity according to the value assigned to an additional sparsity promoting hyperparameter $\lambda$.
The SBL algorithm \cite{tipping2001sparse} proceeds by iteratively updating the estimates of the components of $\w$ and the hyperparameters of the prior using an Expectation-Maximization (EM) algorithm, and by eliminating vectors from the initial function dictionary that are assigned variances below a given threshold. Ultimately, the algorithm selects a sparse subset of terms,  which are assigned Gaussian stochastic parameters. The SBL training is repeated for various fixed values of $\lambda$, and the best-performing model is finally selected after an \emph{a posteriori} cross-validation step: the candidate models are propagated through the CFD solver and the ones providing the lowest predictive error on a set of output Quantities of Interest (QoIs) are finally selected for each class of flows. The detailed SBL-SpaRTA procedure can be found in \cite{cherroud2022sparse}.

\subsection{Flow solver}
The data-driven models resulting from the SBL-SpaRTA algorithm are implemented within a modified version of the open-source finite-volume solver OpenFoam \citep{weller1998tensorial}. The corresponding code can be downloaded from the public github repository:\\ \url{https://github.com/shmlzr/general\_earsm.git}.

In the numerical tests presented below, the governing equations are solved using the well-known SIMPLE algorithm (SimpleFoam library). The linear upwinding is applied to the convective terms, while viscous terms are approximated with a second-order central difference scheme. The solution is advanced to the steady state by using a Gauss-Seidel smoother. For each case, we used the same computational settings and the same grids as in \cite{schmelzer2020discovery}, to which we refer for more details. The grids are fine enough to ensure that the discretization error is negligible compared to the effect of the turbulence model. 

\subsection{Stochastic flow predictions}
The posterior probability distributions of the selected model parameters, along with the associated tensor terms, constitute a stochastic turbulence model correction. The latter can be propagated through the flow solver by using a suitable UQ algorithm, to obtain a stochastic prediction of a new flow. In the following numerical studies, we use the \emph{equadratures} library of \cite{seshadri2017equadratures}. The latter performs sparse regression with a Polynomial Chaos (PC) expansion using again the SBL algorithm to select the most relevant polynomials. The mean and the variance of the decomposition can be then calculated using the sparse coefficients of the PC expansion. The reader is referred to \cite{cherroud2022sparse} for further details. \newline

\begin{figure}[H]
  \centerline{
  \includegraphics[width=0.7\linewidth]{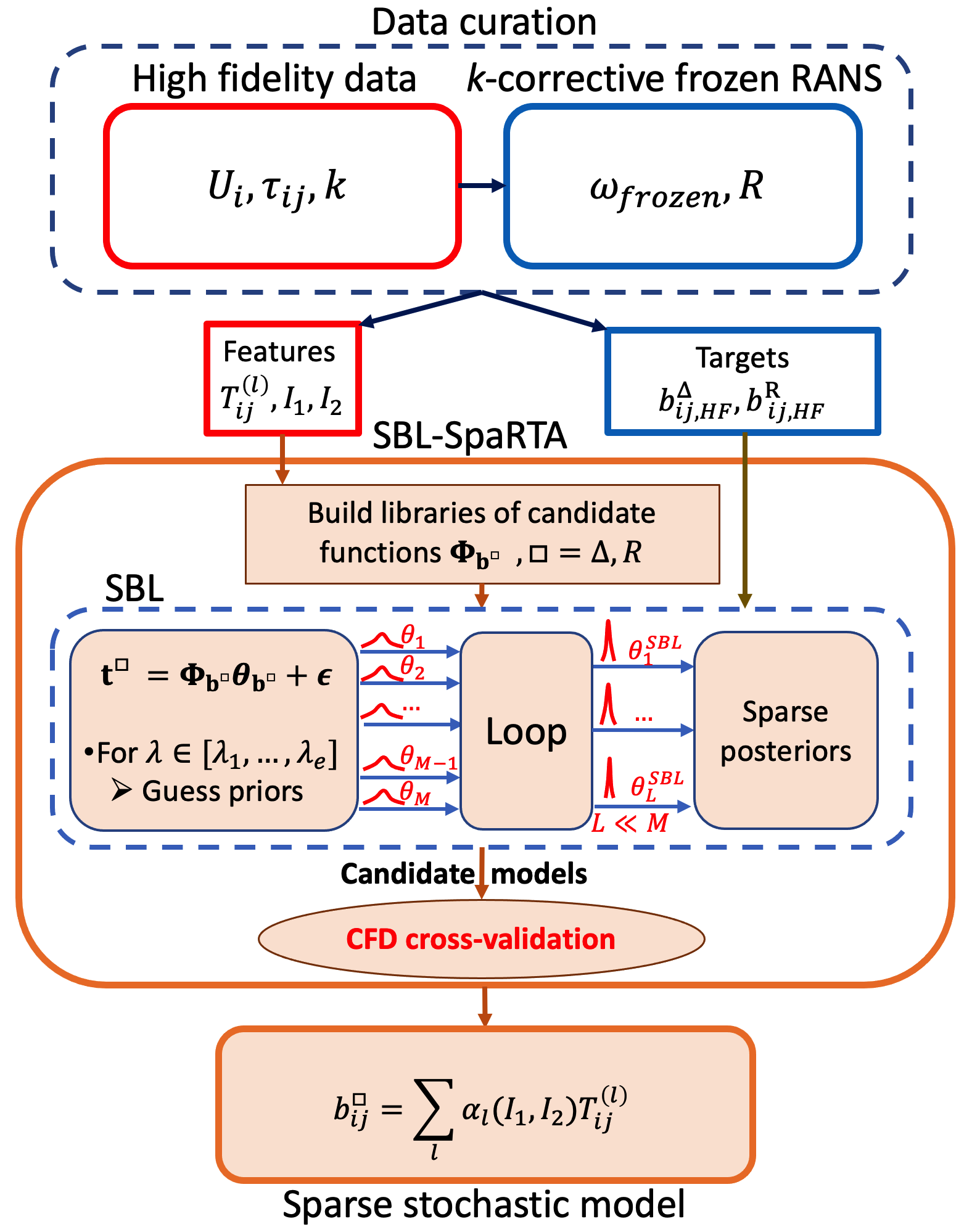}}
  \caption{SBL-SpaRTA workflow.}\label{fig:UQ}
\end{figure}

The complete SBL-SpaRTA workflow, including the UQ procedure, is presented in Figure \ref{fig:UQ}.

\subsection{Space-dependent Model Aggregation}\label{sec:spacedependent}
The customized SBL-SpaRTA models improve the solution of the baseline model for the class of flows for which they are trained, but generally have poor performance for other classes of flows. In addition, it may be difficult to know a priori which model will perform better for an unseen flow that shares common characteristics with different flow classes. \newline

With the goal of providing improved and more generalizable flow predictions while estimating uncertainties associated with the turbulence modeling choices, we introduce a space-dependent model aggregation procedure, referred to as XMA, in which the predictions of multiple models are combined using weighting functions that can vary across the computational domain. Such functions are trained to automatically assign high weights to models that are likely to perform better in a given flow regime or flow region, and low weights to models that are likely to perform poorly.
More specifically, in the following we build on the space-dependent model aggregation approach originally proposed by \cite{MZB2023} for combining the predictions of a set of LEVMs from the literature, and we develop a methodology that locally combines the solutions of a set of competing SBL-SpaRTA models, learned for different flow environments, to predict unseen flows.

\subsubsection{XMA methodology}\label{ext:methodo}
Let us consider $K$ SBL-SpaRTA models, learned in different flow environments, and let $d(\mathbf{x})$ be any Quantity of Interest (QoI) predictable as an output of a RANS flow solver at some spatial location $\mathbf{x}$ (\emph{e.g.} the predicted velocity or pressure fields, the skin friction distribution, etc.). 
In order to make predictions of $d$ that are robust to the choice of the data-driven turbulence model for an unseen flow scenario, we borrow the "Mixture-of-Experts" concept \cite{yuksel_twenty_2012} and we build an ensemble solution by aggregating the individual solutions $d_k$ of the $K$ SBL-SpaRTA models:
\begin{equation}
d_\text{XMA}(\mathbf{x}) = \sum_{k=1}^K w_k(\mathbf{x})d_k(\mathbf{x})
\end{equation}
In the above, $w_k(\mathbf{\mathbf{x}})$ is the weighting function assigned to the $k^{th}$ component model, and $d_\text{XMA}(\mathbf{x})$ is the model ensemble or aggregated prediction. Following the approach of \cite{MZB2023}, we look for weighting functions satisfying the conditions:
\begin{equation}
\begin{dcases}
0\leq w_k(\mathbf{x})\leq 1\quad\forall k=1,...,K \\
\sum_{k=1}^K w_k(\mathbf{x})=1\quad\forall \mathbf{x}
\end{dcases}
\end{equation}
Given the preceding properties, they can be interpreted as the probability of model $k$ to contribute to the aggregated prediction $d_\text{XMA}(\mathbf{x})$. \newline

The weighting functions are constructed as the Exponentially Weighted Average (EWA) of the component model prediction errors:
\begin{equation}\label{Eq:weights}
 w_k\bigl(\delta^{(k)}(\mathbf{x});\bar{\delta}(\mathbf{x});\sigma_{w}\bigr) = \frac{g_k\bigl(\delta^{(k)}(\mathbf{x});\bar{\delta}(\mathbf{x});\sigma_{w}\bigr)}{\DPS \sum_{l=1}^{K} g_l\bigl(\delta^{(l)}(\mathbf{x});\bar{\delta}(\mathbf{x});\sigma_{w}\bigr)}
\end{equation}
where $g_k$ is a gating function of the form
\begin{equation}
g_k\bigl(\delta^{(k)}(\mathbf{x});\bar{\delta}(\mathbf{x});\sigma_{w}\bigr) = \exp \left(- \frac{\Bigl(\delta^{(k)}(\mathbf{x}) - \bar{\delta}(\mathbf{x})\Bigr)^T\,\Bigl(\delta^{(k)}(\mathbf{x}) - \bar{\delta}(\mathbf{x})\Bigr)}{2\sigma_w^2} \right)
\label{Eq:MoE}
\end{equation}
where
\begin{itemize}
\item $\bar{\delta}$ is a vector of high-fidelity data (with $\delta$ being any observed quantity, not necessarily the same as the QoI $d$),
\item $\delta^{(k)}$ is the $k^{th}$ model's output for $\bar{\delta}$,
\item $\sigma_w$ is a hyperparameter.
\end{itemize}
This gating function corresponds to an exponential transformation of the mean square error of the $k^{th}$ model prediction $d_k$ with respect to the high-fidelity data or, put in other terms, to the "score" assigned to the $k^{th}$ model, 
and it can be interpreted as the likelihood of observing the data under some level of observational noise $\sigma_w$.
%
The latter is treated as a hyperparameter, whose role is to discriminate more or less sharply the component models: when $\sigma_w$ is large, all models tend to be assigned approximately equal weights, \ie{} uncertainty on model choice is high, whereas when $\sigma_w$ tends to zero, a single model is selected. Of note, selecting a single model leads to better results if a "true" model is actually included in the multi-model ensemble, but may lead to large errors if none of the models actually captures the truth. The choice of $\sigma_w$ then determines the robustness of the mixture-model prediction against possible over-confident selection of a single model.
In practice, a compromise between robustness and accuracy is needed. In the following numerical experiments, $\sigma_w$ is determined by means of a grid search (see Section \ref{ext-XMA-complete}).\newline

In Equation \eqref{Eq:MoE}, the gating functions and weights are calculated at specific spatial locations $\x$ in the dataset where data are available. Their values at locations outside the dataset are estimated by constructing an approximated regressor $\x\rightarrow w_k(\x)$. Since regression as a function of spatial coordinates would not be applicable to geometries differing from those in the data set, we express instead the regressor as a function a set of local flow features $\boldsymbol{\eta}(\x)$:
\begin{linenomath*}
  \begin{equation*}
    \boldsymbol{\eta}(\mathbf{x}) \rightarrow w_k(\mathbf{x})
  \end{equation*}
\end{linenomath*}
Specifically, we adopt a selection of features proposed in Ref. \cite{ling2015evaluation}, which were successfully used in \cite{MZB2023} for the space-dependent aggregation of turbulence models. The detailed list is given in Table \ref{Table_features}.
\begin{table}[H]
\centering
\centerline{
\resizebox{1.1\textwidth}{!}{%
\begin{tabular}{lccclcc}
  \textbf{Feature} & \textbf{Description} & \textbf{Formula} & & \textbf{Feature} & \textbf{Description} & \textbf{Formula} \\
  \hhline{===~===}
  $\expcomp_1$ & Normalized $Q$ criterion & $\dfrac{||\Ot||^2 - ||\St||^2 }{||\Ot||^2 + ||\St||^2}$ & & $\expcomp_7$ & \muline{Ratio of pressure \\ normal stresses to \\ normal shearstresses} & $\dfrac{\sqrt{\DPS\derp{\pres}{x_i} \derp{\pres}{x_i}}}{\DPS\sqrt{\derp{\pres}{x_j} \derp{\pres}{x_j}} + 0.5\rho\derp{\vel_k^2 }{x_k}}$ \\
  \hhline{---~---}
  $\expcomp_2$ & Turbulence intensity & $\dfrac{k}{0.5\vel_i\vel_i+k}$ & & $\expcomp_8$ & \muline{Non-orthogonality \\ marker between velocity \\ and its gradient \cite{gorle2013framework}} & $\dfrac{\DPS\left|\vel_k\vel_l\derp{\vel_k}{x_l} \right|}{\DPS\sqrt{\vel_n \vel_n\vel_i\derp{\vel_i}{x_j} \vel_m\derp{\vel_m}{x_j}} +\left| \vel_i\vel_j \derp{\vel_i}{x_j} \right|}$ \\
  \hhline{---~---}
  $\expcomp_3$ & \muline{Turbulent Reynolds \\ number} & $\min{\left(\dfrac{\sqrt{k}\lambda}{50 \nu}, 2\right)}$ & & $\expcomp_{9}$ & \muline{Ratio of convection to \\ production of $k$} & $\dfrac{\DPS\vel_i\derp{k}{x_i}}{\DPS|\overline{u_j' u_l'}S_{jl} | + \vel_l \derp{k}{x_l} }$ \\
  \hhline{---~---}
  $\expcomp_4$ & \muline{Pressure gradient \\ along streamline} & $\dfrac{ \DPS\vel_k\derp{\pres}{x_k} }{\DPS \sqrt{\derp{\pres}{x_j}\derp{\pres}{x_j} \vel_i \vel_i} + \left|\vel_l\derp{\pres}{x_l }\right|}$ & & $\expcomp_{10}$ & \muline{Ratio of total Reynolds \\ stresses to normal \\ Reynolds stresses} & $\dfrac{||\overline{u_i' u_j'}||}{ k + ||\overline{u_i' u_j'}|| } $ \\
  \hhline{---~---}
  $\expcomp_5$ & \muline{Ratio of turbulent \\ timescale to mean \\ strain time scale} & $\dfrac{ ||\St|| k}{ ||\St|| k + \varepsilon }$ & & $\expcomp_{11}$ & \muline{Ratio of turbulent \\ production to \\ turbulent dissipation} & $\DPS\frac{P_k}{P_k + \epsilon}$ \\
  \hhline{---~---}
$\expcomp_6$ & Viscosity ratio & $ \dfrac{\nu_T}{100\nu + \nu_T}$  & & &  \\
\hhline{---~~~~}
\end{tabular}%
}}
\caption{ List of input features used to construct the XMA weighting functions.}\label{Table_features}
\end{table}
The selected features are based on domain knowledge, and they include variables such as the $Q$ criterion for vortical flow detection, the turbulent kinetic energy, and the specific rate of dissipation among others. 
The extra feature $ \eta_{11} $, suggested by \cite{girimaji2022NASA}, allows to detect regions where the baseline model already provides reliable solutions and should not be corrected. This is the case of freely decaying turbulence, corresponding to $P_k \rightarrow 0$, \ie{} $ \eta_{11} \rightarrow 0$ or of equilibrium turbulence, where $P_k \rightarrow \epsilon$, which implies $\eta_{11} \rightarrow \frac{1}{2}$. \newline

Random Forests (RF) are used as the regression model; the regression is computed through the \emph{python} package \emph{scikitlearn}\footnote{\url{https://scikit-learn.org}}:
\begin{equation}
\underbrace{\boldsymbol{\eta}(\mathbf{x}) = \bigl(\eta_1(\mathbf{x}),...,\eta_{11}(\mathbf{x})\bigr)}_{\text{local flow features}} \xrightarrow[\mathcal{W}]{RF} 
\underbrace{\left(w_1\bigl(\delta^{(1)}(\mathbf{x});\bar{\delta}(\mathbf{x});\sigma_{w}\bigr),...,w_K \bigl(\delta^{(K)}(\mathbf{x});\bar{\delta}(\mathbf{x});\sigma_{w}\bigr) \right)}_{\text{local model weights}}
\end{equation}
The local flow features are calculated by using the baseline $k$-$\omega$ SST model. 
Ultimately, the XMA model aggregation reads
\begin{equation}\label{xma}
d_\text{XMA}(\mathbf{x}) = \sum_{k=1}^K w_k(\boldsymbol{\eta}(\mathbf{x}))d_k(\mathbf{x}).
\end{equation}
Under the assumption of independent component models, the mean and variance of $d_\text{XMA}(\mathbf{x})$ can be estimated as:
\begin{equation}
 \begin{dcases}
\mathbb{E}\bigl(d_\text{XMA}(\mathbf{x})\bigr) = \sum_{k=1}^K w_k(\boldsymbol{\eta}(\mathbf{x})) \mathbb{E}\bigl(d_k(\mathbf{x})\bigr) \\ 
Var\bigl(d_\text{XMA}(\mathbf{x})\bigr) = \sum_{k=1}^K w_k^2(\boldsymbol{\eta}(\mathbf{x})) Var\bigl(d_k(\mathbf{x})\bigr)
 \end{dcases}
 \label{eq:mean_and_variance}
\end{equation}

\begin{figure}[H]%
\centering
\includegraphics[width=\linewidth]{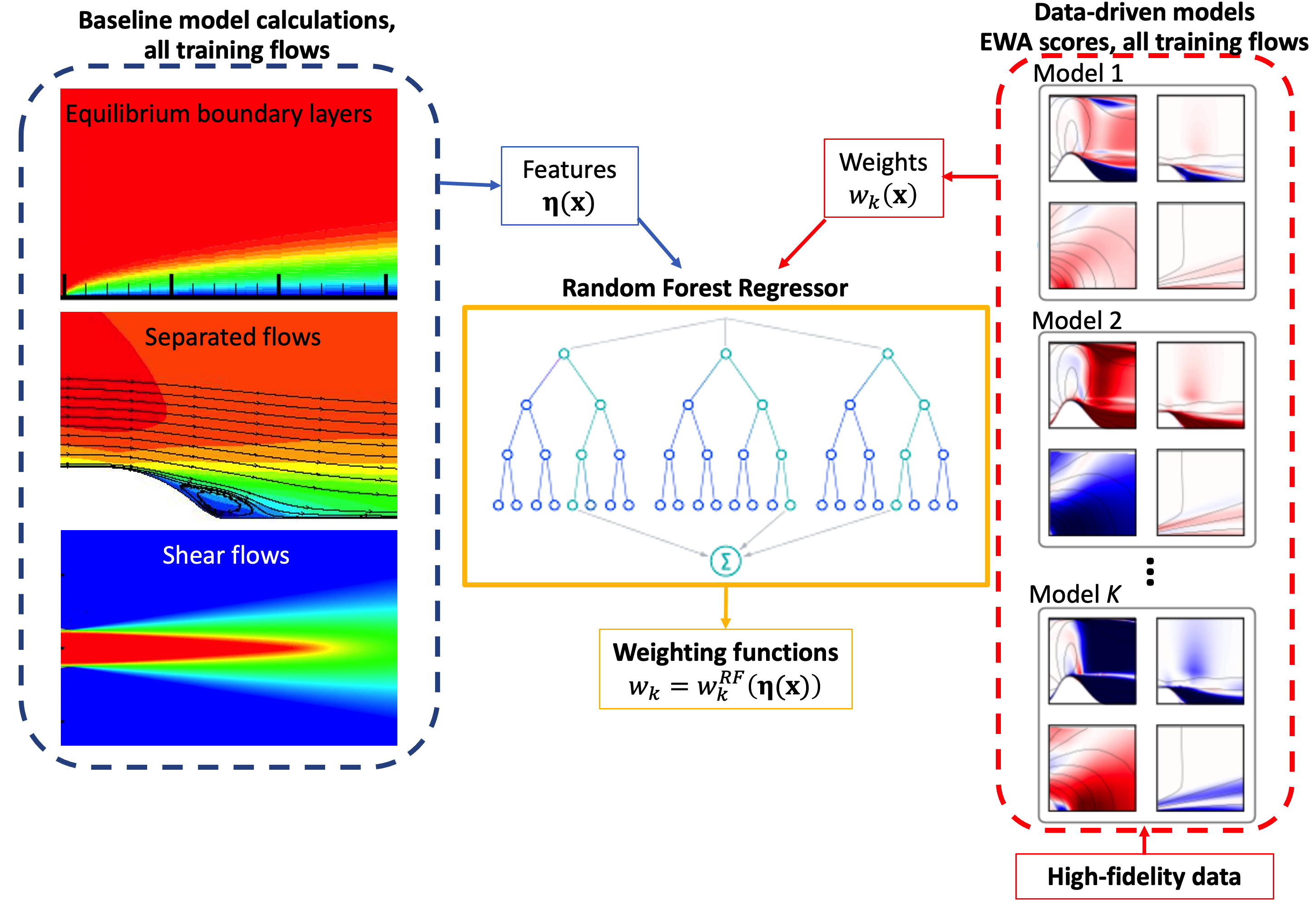}%
\caption[Workflow of XMA training.]{Workflow of XMA training. The baseline $k$-$\omega$ SST model is used to evaluate flow features for a set of training flows including flat plates with various pressure gradients, separated flows and a jet flow (left part); $K$ SBL-SpaRTA model solutions for the training cases are compared with high-fidelity data (right part) to evaluate the gating function (Equation \eqref{Eq:MoE}) and the model weights (Equation \eqref{Eq:weights}). The features and the corresponding weights are used to train Random Forests Regressors that map the local flow features into model weights.}\label{fig:X-MA-workflow}
\end{figure}

The weighting function regression workflow is depicted in Figure \ref{fig:X-MA-workflow}. 

Practical examples of training and application of the data-driven XMA methodology are presented in the next Section.
\section{Model training}\label{modeltraining}
\subsection{Training of the expert models}\label{SBL:modeltraining}
The SBL-SpaRTA algorithm is used to learn stochastic customized models for several classes of flows, listed in Table \ref{tab:training_cases}.
\begin{table}[H]
  \centering
  \begin{tabular}{llcc}
    \multicolumn{2}{l}{\textbf{Training case}} & \textbf{Description} & \textbf{Source} \\
    \hline\hline
    \multicolumn{2}{l}{CHAN} & DNS of incompressible turbulent channel flows & \cite{moser1999direct} \\
               &    & $ Re_{\tau}=180,\,395,\,590$  &  \\
    \hline
    \multicolumn{2}{l}{ZPG} & DNS of an incompressible zero pressure gradient TBL &\cite{schlatter2010assessment} \\
                &   & $ 670,\, 1000,\, 1410,\, 2000,\, 2540,\, 3030,\, 3270,\, 3630,\, 3970,\, 4060$  & \\
    \hline
    \multicolumn{2}{l}{APG} & LES of adverse pressure-gradient TBL & \cite{bobke2017history} \\
                &   & $Re_{\theta} \leq 4000$, $\beta \leq 4$, 5 different pressure gradients & \\
    \hline
    \multicolumn{2}{l}{ANSJ} & PIV of near sonic axisymmetric jet & \cite{bridges2010establishing} \\
    \hline
    SEP & PH & LES of Periodic Hills at $Re=10595$ & \cite{breuer2009flow} \\
    \hhline{~---}
        & CD & DNS of converging-diverging channel at $Re=13600$ & \cite{laval2011direct} \\
    \hhline{~---}
        & CBFS & LES of curved backward facing step at $Re = 13700$ & \cite{bentaleb2012large}\\
        \hline
  \end{tabular}
  \caption{List of flow datasets used to train customized SBL-SpaRTA corrections.}\label{tab:training_cases}
\end{table}
The training cases are representative of diverse physical processes, including incompressible turbulent plane channel flows (CHAN), incompressible turbulent boundary layers (TBL) over flat plates subjected to zero or adverse pressure gradients (ZPG and APG), incompressible separated flows (SEP), and an axisymmetric near sonic jet flow (ANSJ). For each of these flow cases, high-fidelity data corresponding to velocity $\mathbf{U}$, and Reynolds stress tensor $\boldsymbol{\tau}$ are used to construct the target learning vectors ($\tb^{\Delta}$ and $\tb^{R}$) via the corrective frozen RANS procedure, as well as the feature dictionaries ($\PHI_{b^{\Delta}}$ and $\PHI_{R}$). The training is performed using different values of the hyperparameter $\lambda$ within the set $\left\{1,10,10^2,5\times10^2,10^3\right\}$. For each training case, a cross-validation strategy is used to select the sparsity-promoting hyperparameter $\lambda$. To avoid overcharging the training data, models for different $\lambda$ are propagated through the CFD solver, and the models providing the smallest predictive error on the horizontal velocity $U$ are retained as the best ones (see \cite{cherroud2022sparse} for further discussion of the cross-validation step). The model corrections discovered for the various cases after the training and cross-validation steps are presented in Table \ref{tab:learning_models}. They take the general form: 
\begin{equation}
  \M :\;
  \begin{dcases}
    \DPS 
    b_{ij}^{\Delta}=\sum_{l=1}^3\left(\sum_{m,n} \Bigl(\mu_{(m,n)}^{\Delta^{(l)}}\pm\sigma_{m,n}^{\Delta^{(l)}}\Bigr)I_1^m I_2^n
    \right) T_{ij}^{(l)} \pm  \epsilon^{\Delta} \delta_{ij}\\
    \DPS 
    b_{ij}^{R}=\sum_{l=1}^3\left(\sum_{m,n} \Bigl(\mu_{(m,n)}^{R^{(l)}}\pm\sigma_{m,n}^{R^{(l)}}\Bigr)I_1^m I_2^n
    \right) T_{ij}^{(l)} \pm \epsilon^R\delta_{ij} 
  \end{dcases}
\end{equation}
\noindent where $\mu_{(m,n)}^{\Delta^{(l)}}$ and $\sigma_{m,n}^{\Delta^{(l)}}$ (resp. $\mu_{(m,n)}^{R^{(l)}}$ and $\sigma_{m,n}^{R^{(l)}}$) are the mean and the standard deviation, respectively, of the probability density function of the coefficient associated to the term $I_1^m I_2^n$ in the tensor expansion of $b_{ij}^{\Delta}$ (resp. $b_{ij}^{R}$), $\delta_{ij}$ is the Kronecker symbol, and $\epsilon^{\Delta}$ (resp. $\epsilon^R$) is the standard deviation of the noise associated with the model for $b_{ij}^{\Delta}$ (resp. $b_{ij}^{R}$).

\begin{table}[H]
  \centering
  \begin{tabular}{ll}
    \textbf{Expert model}& \quad\textbf{Symbolic expression} \\
    \hline\hline
    $\M^{(CHAN)}$&\;$\,\begin{cases}
      b_{CHAN,ij}^{\Delta} = & [0] \pm 0.0914 \,\delta_{ij}\\
      b_{CHAN,ij}^{R}=& [0] \pm4.61\times10^{-3}\,\delta_{ij}
    \end{cases}$\\
    \hline
    $\M^{(ZPG)}\;$&\;$\;\begin{cases}
      b_{ZPG,ij}^{\Delta}= & [(0.152\pm0.0430)(I_1 - I_2)] T_{ij}^{(1)} \pm 0.167\,\delta_{ij} \\
      b_{ZPG,ij}^{R}=& [0] \pm3.01\times10^{-3}\,\delta_{ij}
    \end{cases}$ \\
    \hline
    $\M^{(APG)}\;$&\;$\;\begin{cases}
      b_{APG,ij}^{\Delta} = & [(2.99\pm0.00726)] T_{ij}^{(2)} \pm 0.000277\,\delta_{ij} \\
      b_{APG,ij}^{R}=& [0] \pm6.55\times10^{-5} \,\delta_{ij}
    \end{cases}$ \\
    \hline
    $\M^{(ANSJ)}$&\;$\;\begin{cases}
     b_{ANSJ,ij}^{\Delta} = & [(0.33\pm0.0189)] T_{ij}^{(1)} \pm 0.00622 \,\delta_{ij}\\
     b_{ANSJ,ij}^{R}=& [0] \pm3.45\times10^{-3}\,\delta_{ij}
    \end{cases}$ \\
    \hline
    $\M^{(SEP)}$&\,$\;\begin{cases}
      b_{SEP,ij}^{\Delta}= & [(5.21 \pm 0.0173)] T_{ij}^{(2)} \pm 0.0348 \,\delta_{ij}\\
      b_{SEP,ij}^{R}=& [(0.681 \pm 0.02)]T_{ij}^{(1)}\pm0.0318\,\delta_{ij}
    \end{cases}$\\
    \hline
  \end{tabular}
  \caption{Customized SBL-SpaRTA corrections obtained for various training flow cases.}\label{tab:learning_models}
\end{table}

A clear advantage of symbolic regression methods, such as SBL-SpaRTA, is that they provide tangible analytical expressions that allow the effects of the discovered model corrections to be interpreted. 
For simplicity, we discuss below the expected behavior of the models of Table \ref{tab:learning_models} by assuming that the stochastic model parameters are fixed to their mean values (i.e. we consider the deterministic results, without uncertainty propagation). Since model posteriors are generally well informed, \emph{i.e.} sharply peaked around their mean, this is sufficient to understand the model main trend (the solution computed at the mean value being a first-order approximation of the mean of the stochastic solution).
The following considerations are in order:
\begin{itemize}
\item $\M^{(CHAN)}$: the optimal correction for the channel flow model is zero to within observation noise, confirming that the baseline model is well fit for this canonical flow, at least as long as the main goal is to provide a good prediction of the velocity field, since the shear Reynolds stress largely dominates the other stresses. Given the great similarity of the CHAN model with the baseline $k-\omega$ model (noted hereafter $\M^{(BL)}$, where $BL$ stands for baseline), we do no longer consider it in the following analyses.
\item $\M^{(ZPG)}$: the model discovered using zero pressure gradient boundary layer data exhibits a small anisotropy correction dependent on the linear term $T_{ij}^{(1)}=S_{ij}/\omega$. Such correction adds to the Boussinesq term and leads to the corrected anisotropy relation:
  \begin{linenomath*}
    \begin{equation*}
      2kb_{ZPG,ij}\DIFaddbegin \DIFadd{^{\Delta}}\DIFaddend =\left[-2\nu_t+ 0.152(I_1 - I_2)\frac{2 k}{\omega}\right]S_{ij}=
      -\frac{2k}{\omega}\alpha_{ZPG}S_{ij},\quad \alpha_{ZPG}\approx 1-0.152(I_1 - I_2)
    \end{equation*}
  \end{linenomath*}
  The coefficient $\alpha_{ZPG}$ corresponds to a very small correction (decrease) of eddy viscosity $\nu_t$ in the external region, as shown in Figure \ref{fig:channut} for one of the CHAN flow cases ($Re_\tau=1000$). The slope of the velocity profiles in log layer remain essentially unchanged with respect to the baseline (Figure \ref{fig:chanu+}). Note that, for incompressible boundary layer flows, $I_1\approx - I_2\approx \displaystyle\frac{1}{2\omega^2}\left(\frac{\partial U}{\partial y}\right)^2$, meaning that the correction is active when the shear timescale is smaller than the turbulent timescale. On the other hand, no additional corrective term is added to the turbulent kinetic energy equation on average, except for the small negative contribution of $b_{ij}^\Delta$ to the production term (Eq. \eqref{datacuration:production}). This indicates that the discovered optimal model for the ZPG boundary layer case is also very close to the baseline, which is in agreement with our \emph{a priori} belief that $k-\omega$ SST has been well tuned for these kinds of flows.
  \begin{figure}[H]
    \centering
    \begin{subfigure}[b]{0.49\linewidth}
      \centering
      \includegraphics[width=\linewidth]{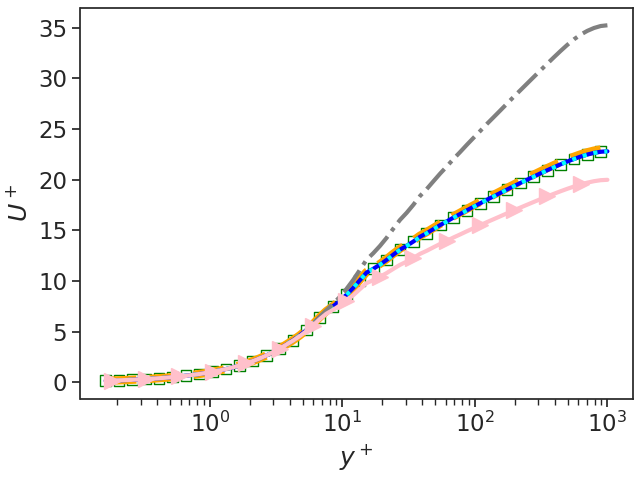}
      \caption{Horizontal velocity profile}
      \label{fig:chanu+}
    \end{subfigure}
    \hfill
    \begin{subfigure}[b]{0.49\linewidth}
      \centering
      \includegraphics[width=\linewidth]{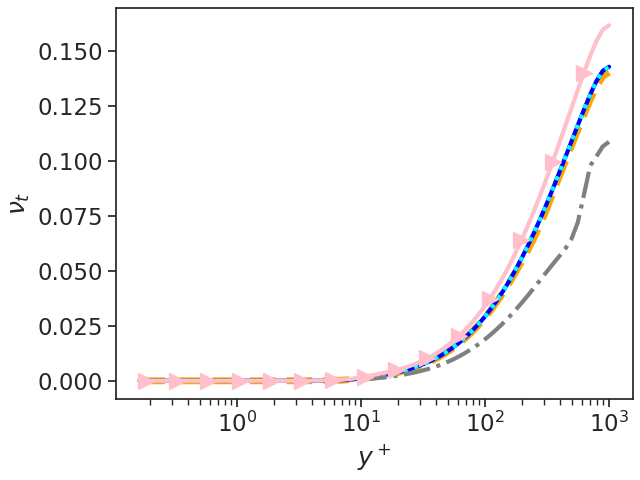}
      \caption{Eddy viscosity $\nu_t$ profile.}
      \label{fig:channut}
    \end{subfigure}
    \caption{Incompressible turbulent channel flow at $Re_{\tau}=1000$: wall-normal profiles of the streamwise velocity (in wall units) $U^+$ and of the eddy viscosity $\nu_t$ as a function of the non-dimensional wall normal direction $y^+=y U_{\tau}/\nu$. - $\M^{(ZPG)}$ (\dashed), $\M^{(CHAN)}=\M^{(BL)}$ (\solid), $\M^{(APG)}$ (\dotted), $\M^{(ANSJ)}$ (\chain), $\M^{(SEP)}$ (\triangle) and high-fidelity data (\square).}
    \label{fig:CHAN_plots}
  \end{figure}
\item $\M^{(APG)}$: the model discovered for the adverse pressure gradient boundary layer cases contains a nonlinear anisotropy correction, resulting in a constitutive relation of the form:
  \begin{linenomath*}
    \begin{equation*}
      2k \,b_{APG,ij}\DIFaddbegin \DIFadd{^{\Delta} }\DIFaddend = -2\nu_t S_{ij}+2.99\frac{2k}{\omega^2}\left[S_{ik}\Omega_{kj}-\Omega_{ik}S_{kj}\right]
    \end{equation*}
  \end{linenomath*}
  Here again no additional corrective term is selected for the turbulent kinetic energy equation, and the contribution of $b_{ij}^\Delta$ to the TKE production is zero because the tensor product of $\T^{(2)}$ with the velocity gradient tensor is null. This is confirmed by inspection of Figure \ref{fig:channut}, where the eddy viscosity profile across the CHAN flow predicted by the APG model is superimposed with those of the baseline model. The correction also has essentially no effect on the velocity profile for both the channel flow case shown in Figure \ref{fig:chanu+} and the ANSJ case (Figure \ref{fig:ANSJUaxistraining}), and it provides skin friction profiles of separated cases such as the CD and PH flows in close agreement with the baseline model (see Figure \ref{fig:SEP_plots}), showing that the learned correction for APG plays a very minor role for a large variety of cases. 
\begin{figure}[H]
  \centering
  \begin{subfigure}[b]{0.49\linewidth}
    \centering
    \includegraphics[width=\linewidth]{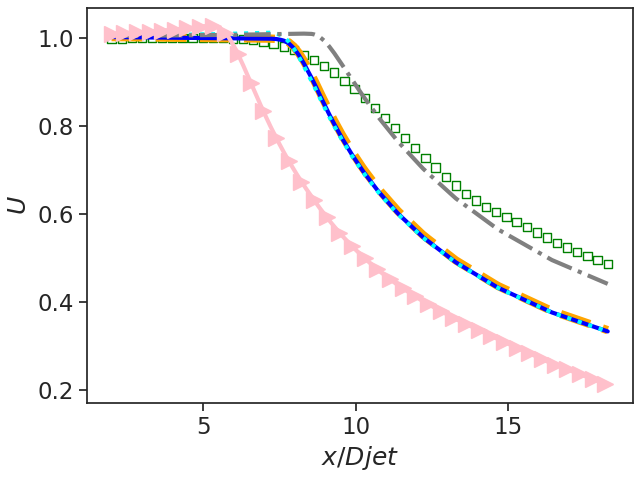}
    \caption{Horizontal velocity $U$.}
    \label{fig:ANSJUaxistraining}
  \end{subfigure}
  \hfill
  \begin{subfigure}[b]{0.49\linewidth}
    \centering
    \includegraphics[width=\linewidth]{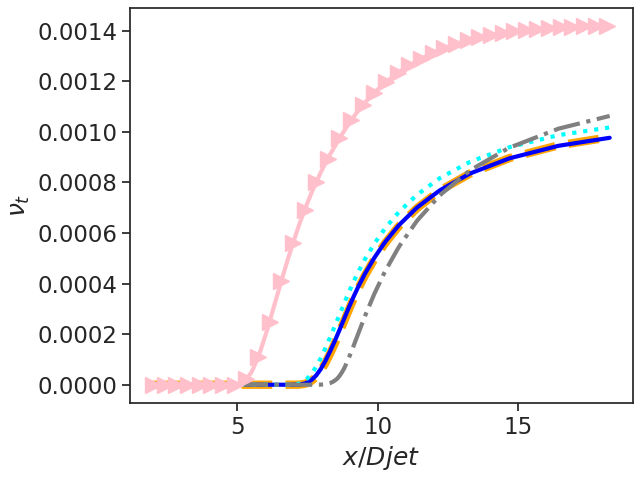}
    \caption{Eddy viscosity $\nu_t$}
    \label{fig:ANSJnuttraining}
  \end{subfigure}
  \caption{Axisymmetric near-sonic jet flow: distributions of the horizontal velocity (a) and of the eddy viscosity (b) along the jet axis as a function of the non-dimensional streamwise coordinate $x/D_{jet}$ (with $D_{jet}$ the diameter of the jet's exhaust). - $\M^{(ZPG)}$ (\dashed), $\M^{(CHAN)}=\M^{(BL)}$ (\solid), $\M^{(APG)}$ (\dotted), $\M^{(ANSJ)}$ (\chain), $\M^{(SEP)}$ (\triangle) and High-fidelity data (\square).}
  \label{fig:ANSJ_plots}
\end{figure}
\begin{figure}[H]
  \centering
  \begin{subfigure}[b]{0.49\linewidth}
    \centering
    \includegraphics[width=\linewidth]{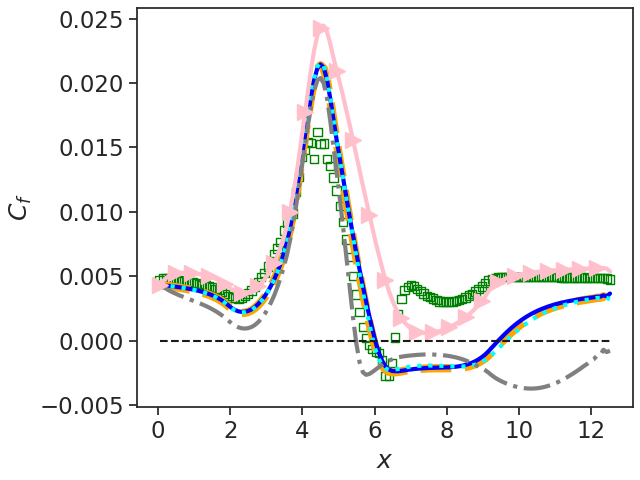}
    \caption{Separated converging-diverging channel flow (CD).}
    \label{fig:CDCf}
  \end{subfigure}
  \hfill
  \begin{subfigure}[b]{0.49\linewidth}
    \centering
    \includegraphics[width=\linewidth]{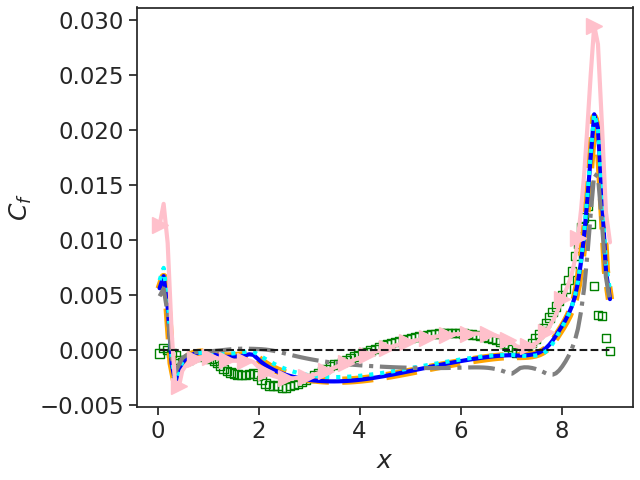}
    \caption{Periodic hill flow (PH).}
    \label{fig:PHCf}
  \end{subfigure}
  \caption{Skin-friction $C_f$ distribution along the bottom wall as a function of the streamwise coordinate. $x$ - $\M^{(ZPG)}$ (\dashed), $\M^{(CHAN)}=\M^{(BL)}$ (\solid), $\M^{(APG)}$ (\dotted), $\M^{(ANSJ)}$ (\chain), $\M^{(SEP)}$ (\triangle) and High-fidelity data (\square).}\label{fig:SEP_plots}
\end{figure}
\item $\M^{(ANSJ)}$: the model consists of a linear anisotropy correction, resulting in a decreased eddy viscosity compared to the baseline:
  \begin{linenomath*}
    \begin{equation*}
      2k\,b_{ANSJ,ij}\DIFaddbegin \DIFadd{^{\Delta} }\DIFaddend =\left[-2\nu_t+0.33\frac{2 k}{\omega}\right]S_{ij} = -\frac{2k}{\omega}\alpha_{ANSJ}S_{ij},\quad \alpha_{ANSJ}\approx 0.67
    \end{equation*}
  \end{linenomath*}
  Such a modification improves significantly the model accuracy for the training flow case, and specifically the jet spreading rate, as shown in Figure \ref{fig:ANSJUaxistraining}. However, this leads to a severe underestimation of the eddy viscosity for the channel flow case (Figure \ref{fig:ANSJnuttraining}) and consequently of the wall friction, resulting in a wrong log-law slope, as shown in Figures \ref{fig:chanu+} and \ref{fig:channut}. This also results in an underestimated skin friction and in a too large separation bubble for both CD and PH flows, reported in Figures \ref{fig:CDCf} and \ref{fig:PHCf}. 
\item $\M^{(SEP)}$: the model trained on the separated flow cases involves both a nonlinear correction to the extra anisotropy and a linear $\bb^R$ correction.
  \begin{linenomath*}
    \begin{equation*}
      \begin{dcases}{}
        2k\,b_{SEP,ij}\DIFaddbegin \DIFadd{^{\Delta} }\DIFaddend = -2\nu_t S_{ij} +5.21\frac{2k}{\omega^2}\left[S_{ik} \Omega_{kj}-\Omega_{ik}  S_{kj} \right] \\
        2k\,b_{SEP,ij}^R =-\frac{2k}{\omega}\alpha_{SEP}S_{ij} ,\quad \alpha_{SEP}\approx 1.181
      \end{dcases}
    \end{equation*}
  \end{linenomath*}
  While the non-linear correction $b_{ij}^{\Delta}$ does not affect the TKE production, the $b_{ij}^R$ correction tends to increase the eddy viscosity. Inspection of the skin friction distribution for the CD and PH flows (Figures \ref{fig:CDCf} and \ref{fig:PHCf}) shows that the SEP model significantly improves the agreement with the high-fidelity data in terms of size and position of the recirculation bubble in the PH case (Figure \ref{fig:PHCf}) compared to all other models, and it also results in a more satisfactory overall agreement for the CD case, but it misses the small separation bubble in divergent (Figure \ref{fig:CDCf}). The reason is that the model overcorrects the baseline, resulting for instance in an overdissipation for the channel flow case (see Figure \ref{fig:channut}) and in an underestimation of the log-law slope (Figure \ref{fig:chanu+}). Furthermore, the SEP model provides inaccurate predictions of the velocity distribution for the ANSJ case (Figure \ref{fig:ANSJUaxistraining}).
\end{itemize}

The preceding discussion shows that the SBL-SpaRTA models trained on a class of flows do not generalize well to flows for which they were not trained. This is why the discovered models are qualified of "customized" models and do not constitute a universal general-purpose model. 

Since complex flows of practical interest may simultaneously exhibit various physical processes, some blending of the customized models is needed. In addition, estimates of the model uncertainties are required to detect whether the models can be trusted when predicting a new unseen flow. These aspects are dealt with by means of the proposed XMA approach, as discussed in the next Section.

\subsection{Training of the XMA weighting functions}\label{ext-XMA-complete}
Once the training of expert models is completed, we subsequently train the weighting functions of the aggregated model.
For that purpose, we consider a set of five component models $\mathcal{K}=\left\{\M^{(BL)}, \M^{(ZPG)},\M^{(APG)}, \M^{(ANSJ)}, \M^{(SEP)} \right\}$ and high-fidelity datasets among those of Table \ref{tab:training_cases}:
$\mathcal{D}=\left\{\mathcal{D}_{CHAN}, \mathcal{D}_{APG}, \mathcal{D}_{ANSJ}, \mathcal{D}_{SEP}\right\}$. 
From such datasets, we extract all streamwise velocity data. Of note, such data were used to generate the high-fidelity features of expert models via the $k$-corrective frozen RANS procedure, but were not used as the training target. 
The XMA training then proceeds as follows:
\begin{enumerate}
\item {\bf Feature evaluation}:  baseline model solutions are computed for all flow configurations in $\mathcal{D}$, and the results are postprocessed to extract the features of Table \ref{Table_features}.
\item {\bf Target data preparation}: the same flows are also simulated with all other models in $\mathcal{K}$, and the velocities are extracted at all locations where high-fidelity data are available.
\item {\bf Hyperparameter tuning}: several candidate values of the hyperparameter $\sigma_w$ are considered, with $2\sigma_w^2  \in \{1,10^{-1},10^{-2},10^{-3},10^{-4}\}$.
\item {\bf Exponentially weighted average}: for each model, each flow case, and each $\sigma_w$ we compute the EWA from Eqs \eqref{Eq:weights}  and \eqref{Eq:MoE}, thus generating target data for the model weights. Such weights are associated to the features computed at the same location, thus constituting a labelled data set: (Features, Weights).
\item  {\bf Random Forest Regressor}: the target data for all flow cases are randomly split in a training and a validation set (of equal size). The latter is used for tuning the EWA hyperparameter $\sigma_w$.
A Random Forest regressor is trained on the training set data and used to infer the weights at the validation data locations. An aggregated velocity estimate is then computed at these location as an XMA of the component models, and finally an "improvement" score of the XMA prediction over the baseline model is measured as:
    \begin{equation}\label{Eq:Imp}
    Imp(\sigma_w)  = \left(  1 -  \frac{ \DPS  ||U^{(XMA)}(\mathbf{x};\sigma_w) - U^{(HF)}(\mathbf{x})||_2^2   }{    \DPS    ||U^{(BL)}(\mathbf{x}) - U^{(HF)}(\mathbf{x})||_2^2} \right) \times 100
    \end{equation}
    The procedure is repeated for all the preselected values of $\sigma_w$.
 \item{\bf Best model selection}: the best-performing regressor is finally selected as the one providing the highest improvement score.
 \end{enumerate}

\begin{figure}[H]
  \centering

  \begin{subfigure}[b]{0.49\linewidth}
    \includegraphics[width=1.\linewidth, trim=35 0 50 0, clip]{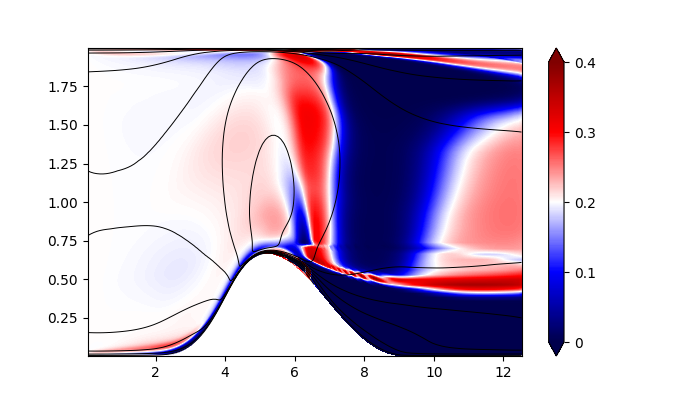}
    \caption{$w_\text{ZPG}$}
  \end{subfigure} \hfill
  \begin{subfigure}[b]{0.49\linewidth}
    \includegraphics[width=1.\linewidth, trim=35 0 50 0, clip]{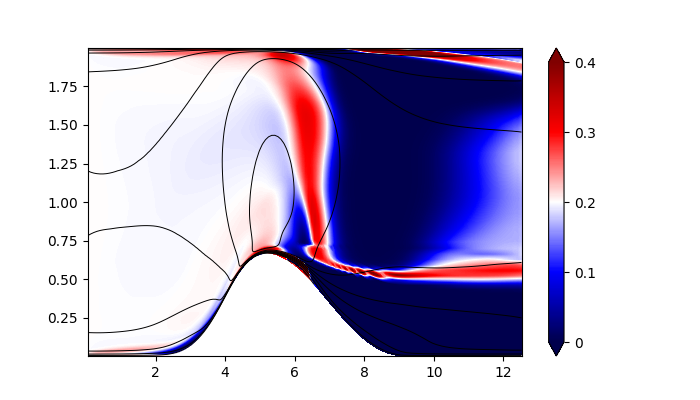}
    \caption{$w_\text{APG}$}
  \end{subfigure}
  \\
  \begin{subfigure}[b]{0.49\linewidth}
    \includegraphics[width=1.\linewidth, trim=35 0 50 0, clip]{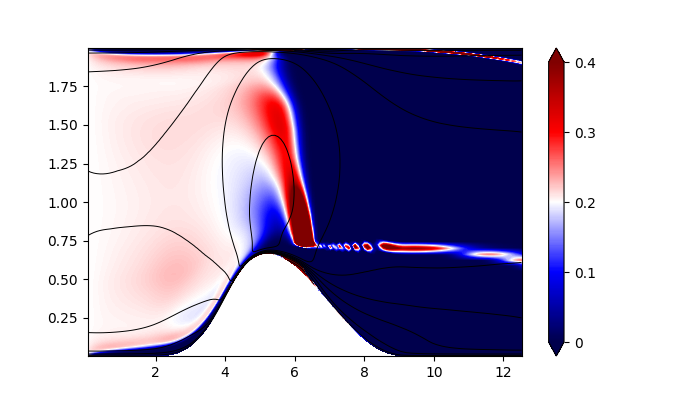}
    \caption{$w_\text{ANSJ}$}
  \end{subfigure} \hfill
  \begin{subfigure}[b]{0.49\linewidth}
    \includegraphics[width=1.\linewidth, trim=35 0 50 0, clip]{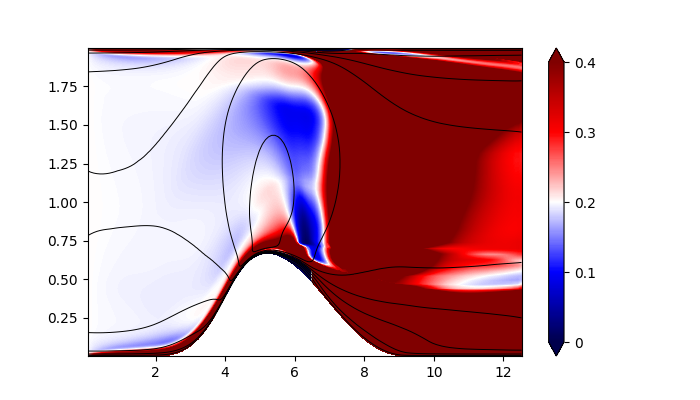}
    \caption{$w_\text{SEP}$}
  \end{subfigure}
  \\
  \centering
\begin{subfigure}[b]{0.49\linewidth}
    \includegraphics[width=1.\linewidth, trim=35 0 50 0, clip]{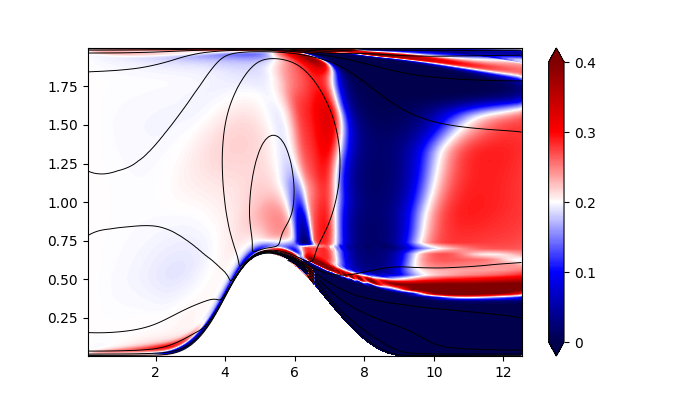}
    \caption{$w_\text{BL}$}
  \end{subfigure}
  \caption{Colormaps of exact model weights for CD and iso-contours of longitudinal velocity (baseline model)\label{fig:wCD}.}\label{fig:enter-label}
\end{figure}

\begin{figure}[H]
  \centering
  \begin{subfigure}[b]{0.49\linewidth}
    \includegraphics[width=1.\linewidth, trim=35 0 50 0, clip]{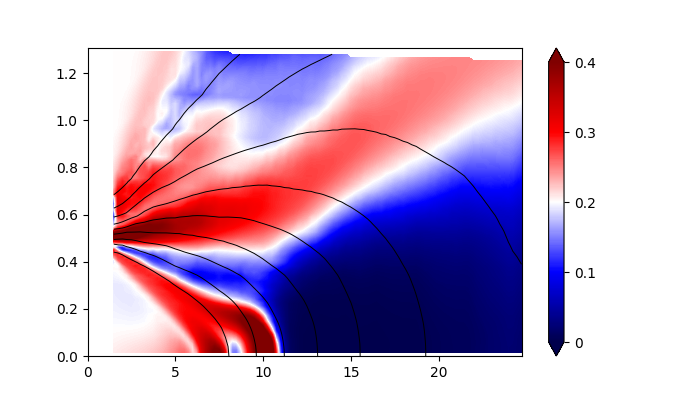}
    \caption{$w_\text{ZPG}$}
  \end{subfigure}
  \hfill
  \begin{subfigure}[b]{0.49\linewidth}
    \includegraphics[width=1.\linewidth, trim=35 0 50 0, clip]{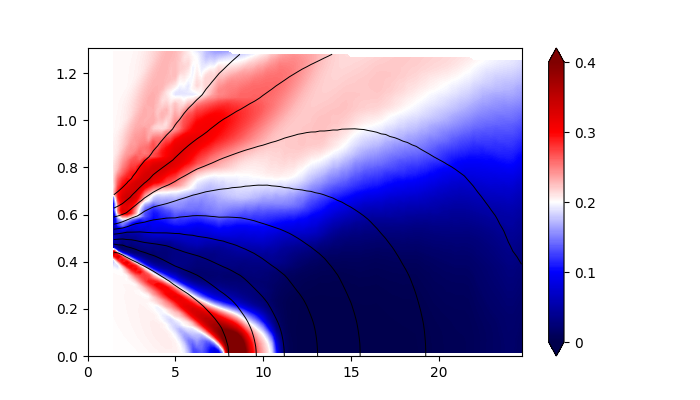}
    \caption{$w_\text{APG}$}
  \end{subfigure}
  \\
  \begin{subfigure}[b]{0.49\linewidth}
    \includegraphics[width=1.\linewidth, trim=35 0 50 0, clip]{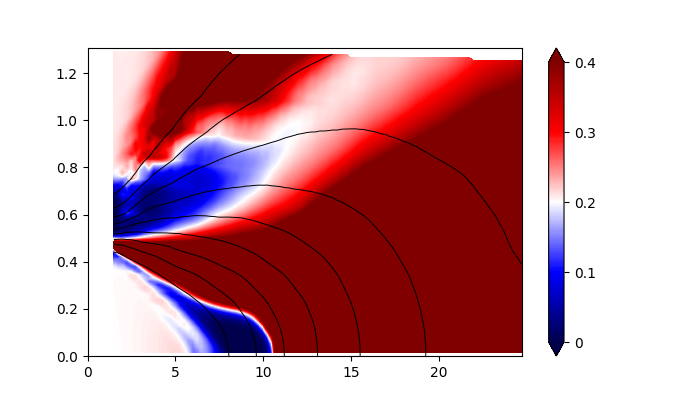}
    \caption{$w_\text{ANSJ}$}
  \end{subfigure} \hfill
  \begin{subfigure}[b]{0.49\linewidth}
    \includegraphics[width=1.\linewidth, trim=35 0 50 0, clip]{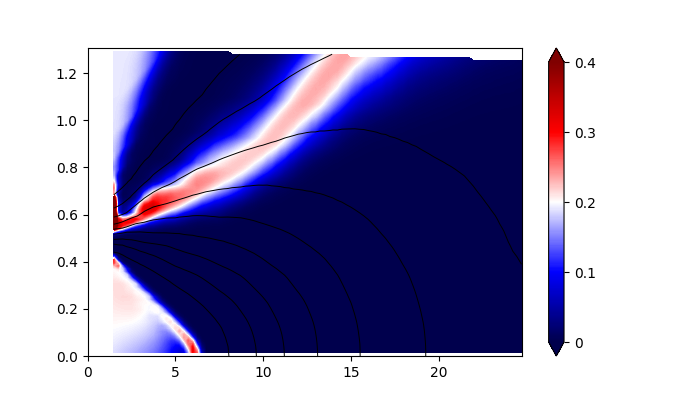}
    \caption{$w_\text{SEP}$}
  \end{subfigure}
  \\
  \centering
\begin{subfigure}[b]{0.49\linewidth}
    \includegraphics[width=1.\linewidth, trim=35 0 50 0, clip]{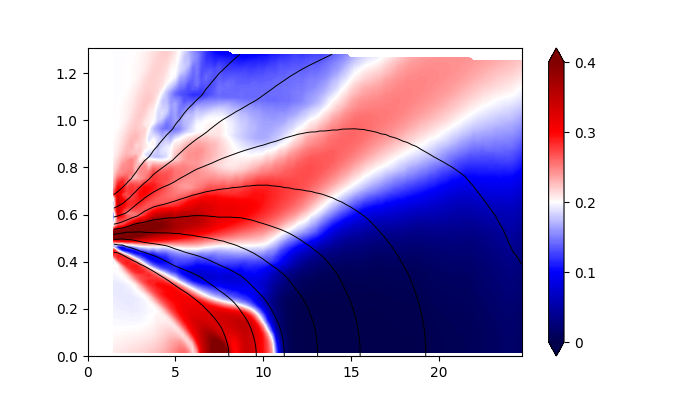}
    \caption{$w_\text{BL}$}
  \end{subfigure} 
 \caption{Colormaps of exact model weights for ANSJ and iso-contours of longitudinal velocity (baseline model)\label{fig:wANSJ}.}
  \label{fig:enter-label}
\end{figure}

Figures \ref{fig:wCD} and \ref{fig:wANSJ} display colormaps of the target weights $w_k$ computed from high-fidelity data for two training cases (namely, CD and ANSJ) and five models, using the tuned $\sigma_w$. In the same figures we also report the iso-contours of the longitudinal velocity of the baseline model, to illustrate how the models are scored in different flow regions. 
In the CD case (Figure \ref{fig:wCD}), all models are scored equally upstream of the throat, except at the beginning of the convergent, where the ANSJ model is downgraded compared to the others. On the other hand, in the separated region downstream of the throat, the SEP model is assigned a much higher weight than the other ones. The ANSJ model exhibits the worst score almost everywhere.
For the ANSJ case (Figure \ref{fig:wANSJ}), the models have essentially equal weights in the potential cone region, which is insensitive to the turbulence model choice. Immediately downstream of this region, the ZPG and APG models exhibit relatively high performance scores. As expected, the ANSJ model is assigned the highest score in the far jet region. On the contrary, the SEP model exhibits a very low score almost everywhere. 

Overall, all of these results highlight that the expert models exhibit a clear regional performance, with each model generally obtaining higher scores in flow regions dominated by physical processes similar to those observed in the model training set. 
The reliability of the Random Forest regressor has been evaluated by means of the coefficient of determination $R^2$,  defined by:
\begin{linenomath*}
  \begin{equation*}
    R^2=1-\frac{\DPS\sum_i (\overline{w}(\x_i)-w(\x_i))^2}{\DPS\sum_i (\overline{w}(\x_i)-\mathbb{E}(\overline{w}(\x_i))^2},\;
    \begin{cases}
      \overline{w}(\x_i):\; \text{computed from high-fidelity data} \\
      w(\x_i):\;\text{RFR predicted}
    \end{cases}
  \end{equation*}
\end{linenomath*}
$R^2$ is greater than 0.97 for both the training set (training error) and the validation set. \newline
The regressor reliability has also been tested on unseen flow configurations, given in Table \ref{unseencases}, extracted from the NASA 2022 Symposium on Turbulence modeling Collaborative Testing Challenge \cite{rumsey2022nasa}.
\begin{table}[H]
  \centering
  \begin{tabular}{lcc}
    \textbf{Case} & \textbf{Description} & \textbf{Source} \\
    \hline\hline
    2DZP & Turbulent flat plate & \href{https://turbmodels.larc.nasa.gov/flatplate.html}{2DZP NASA web page}\\
    ASJ & Axisymetric subsonic jet & \href{https://turbmodels.larc.nasa.gov/jetsubsonic_val.html}{ASJ NASA web page} \\
    2DWMH & 2D Wall-Mounted Hump & \href{https://turbmodels.larc.nasa.gov/nasahump_val.html}{2DWMH NASA web page} \\
    \hline
  \end{tabular}
  \caption{Unseen scenarios. The scenarios presented here are never involved in any learning process, neither for the SBL-SpaRTA models nor for the weights regressions.}\label{unseencases}
\end{table}
For 2DZP and ASJ, the coefficient of determination is greater than 0.95. For 2DWMH, it is greater than 0.85. 
This gives us confidence that the learned weighting functions can be extrapolated to flow cases not included in the training set.
\section{Application of XMA to selected flows}\label{sec:resultsxma}
\begin{figure}[H]%
\centering
\includegraphics[width=\linewidth]{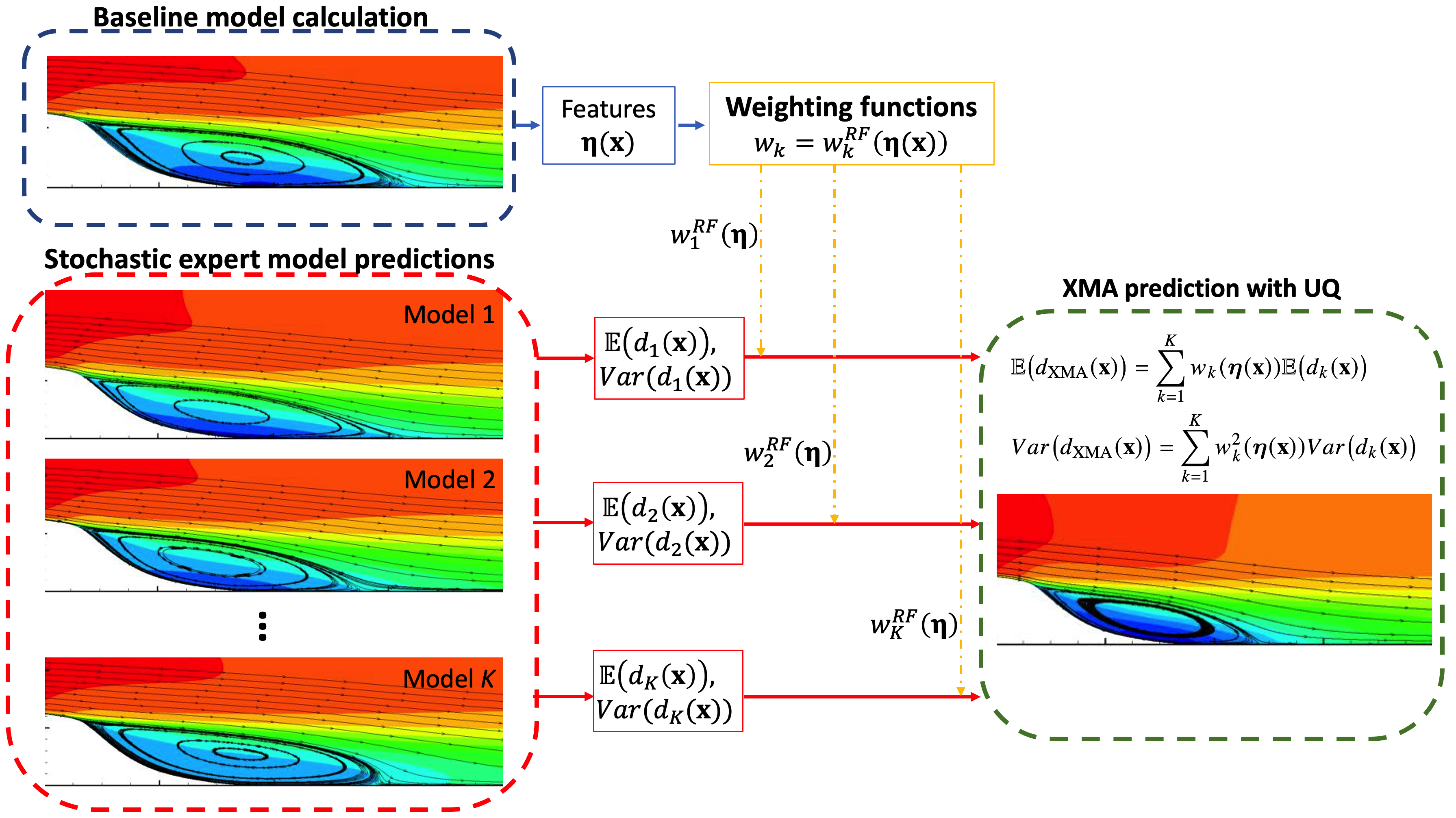}%
\caption[Workflow of XMA predictions.]{Workflow of XMA predictions. Top row: features are extracted from a baseline $k$-$\omega$ SST computations of some new flow; the latter are fed to the RF regressor to obtain weights for the expert models. Bottom row: $K$ stochastic solutions are also obtained for the expert data-driven models by using a suitable UQ method;
the means and variances of the stochastic solutions are aggregated via the weighting function, and the XMA mean and variance are obtained.}\label{fig:X-MA-workflow1}
\end{figure}

In the following, we first apply the XMA approach to two flows included in the training sets used to learn the SBL-SpaRTA corrections or the XMA weighting functions. Then, we evaluate XMA for generalization and compute an aggregated prediction for three unseen flows selected from those proposed in the NASA turbulence modeling testing challenge. \cite{rumsey2022nasa}. In all cases, the XMA results are obtained as follows:
\begin{enumerate}
\item The five models of Table \ref{tab:learning_models} are used to make stochastic predictions of the  flow at stake, \ie{} to estimate the mean $\mathbb{E}(d_k)$ and the variance $Var(d_k)$ of any output flow quantity. At this stage, the parametric uncertainty associated with the learned model parameters is quantified.
\item At each mesh point, the baseline model is used to compute the features of Table \ref{Table_features}.
\item The features are fed to the trained RF to obtain the weighting functions $w_k$.
\item The mean and variance of the XMA aggregated solution are then computed. The XMA variance accounts for both parametric and model-form uncertainty.
\end{enumerate}
Figure \ref{fig:X-MA-workflow1} provides the workflow of XMA stochastic predictions.

\subsection{Application of XMA to flows in the training set}
The XMA is first applied to two flows among those used to train the weighting functions.
We select one of the canonical fully-developed channel flows (CHAN) at $Re_{\tau}=1000$ and the CD separated flow (see Table \ref{tab:training_cases}).

\begin{figure}[H]
  \centering
  \begin{subfigure}{0.49\linewidth}
    \centering
    \includegraphics[width=1.\linewidth, trim=0 0 42 0, clip]{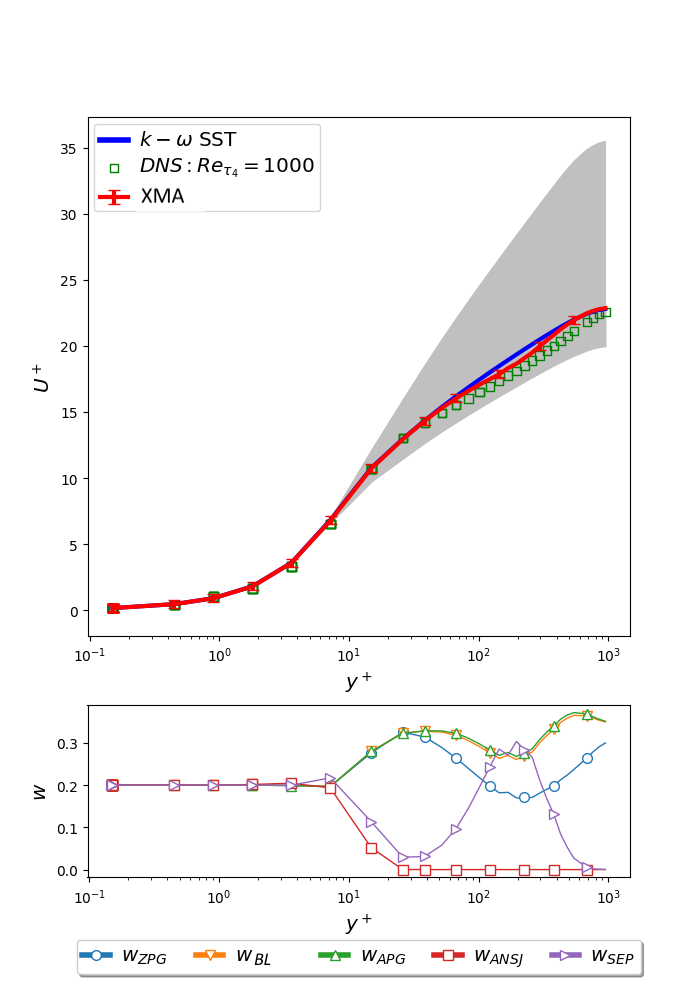}
    \caption{CHAN, $Re_{\tau}=1000$. Horizontal velocity (in wall units) $U^+$ versus the non-dimensional normal coordinate $y^+$.}\label{fig:upluschanw}
  \end{subfigure}
  \hfill
  \begin{subfigure}{0.49\linewidth}
    \centering
    \includegraphics[width=1.\linewidth, trim=0 0 42 0, clip]{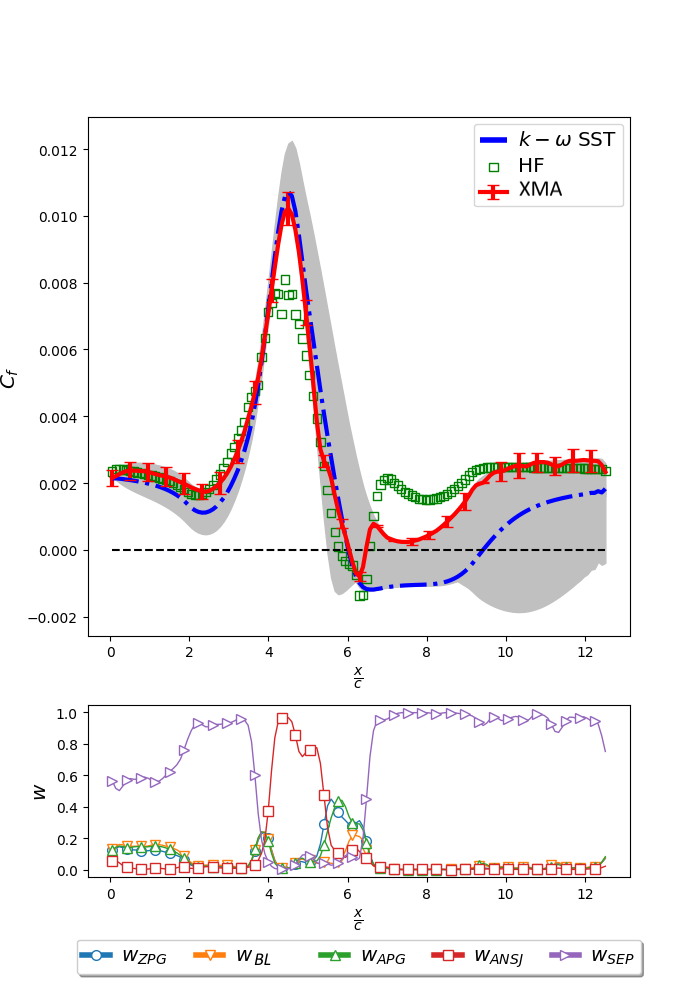}
    \caption{CD case. Skin friction $C_f$ distribution along the bottom wall as a function of the non-dimensional streamwise coordinate $x/c$.}\label{fig:cfsepw}
  \end{subfigure}
  \caption{XMA stochastic predictions for a turbulent channel flow (left) and the converging-diverging channel flow (right). The grey shades represent the accessible region. The error bars correspond to $\pm$ three standard deviations. Distributions of the local model weights are provided in the bottom figures. }
\end{figure}

In Figure \ref{fig:upluschanw} we plot the expectancy of the XMA velocity profile along with the reference high-fidelity data and the baseline model solution. In the picture, the grey-shaded area represents the \emph{convex accessible region}, \ie{} the envelope of solutions given by the five component models. We also reported error bars, corresponding to $\pm 3 \sqrt{Var_\text{XMA}}.$  
The solution is in good agreement with the reference, and it slightly outperforms the baseline $k$-$\omega$ SST model in the log layer. The weights attributed to the various models are reported in the lower panel of the same figure. In the viscous sublayer all models are assigned equal weights, since all models predict the linear solution. In the log layer, the ANSJ model is downgraded, whereas the ZPG and APG models are assigned similar weights because they exhibit a similar performance (see Figure \ref{fig:chanu+}). The SEP model is eventually assigned an increased weight in the defect layer but its contribution remains small overall. Despite the large spread of the accessible area, the solution variance is rather small. This is due to the fact that 1) most models (except ANSJ) predict similar solutions with small posterior variance due to the residual uncertainties in model parameters and 2) the outlier model ANSJ is affected zero weight in the region where it strongly deviates from the other models.

Figure \ref{fig:cfsepw} displays the expectancy and error bars of the friction coefficient distribution for the CD case. For this case, the SEP model is assigned the highest weight (close to 1 in most regions) but the ANSJ model takes over in the throat region where the SEP model is overly dissipative. Of note, the APG model is assigned a slightly higher weight than the other models in the divergent, \ie{} in the adverse pressure gradient region, but its contribution remains similar to the ZPG and CHAN models overall. This confirms that the ZPG, CHAN and APG models behave rather similarly to each other and to the baseline $k$-$\omega$ SST. The automatic selection of the locally best performing models by means of the weighting functions allows to capture the recirculation bubble, which was missed or overestimated by the component models. In most regions, the error bars are small because a single model tends to prevail. The XMA prediction couldn't reach the high-fidelity values of $C_f$ directly after reattachment since the convex accessible envelop of the solutions does not comprise these data points. 

\subsection{XMA prediction of unseen flows}
Next, we assess the XMA for flow cases that were not used for training the SBL-SpaRTA models or the RF weighting functions, namely, the three cases of Table \ref{unseencases}. 

\subsubsection{Turbulent flat plate (2DZP)}
\begin{figure}[H]
  \begin{subfigure}[b]{0.49\linewidth}
    \includegraphics[width=1.0\linewidth, trim=0 0 42 0, clip]{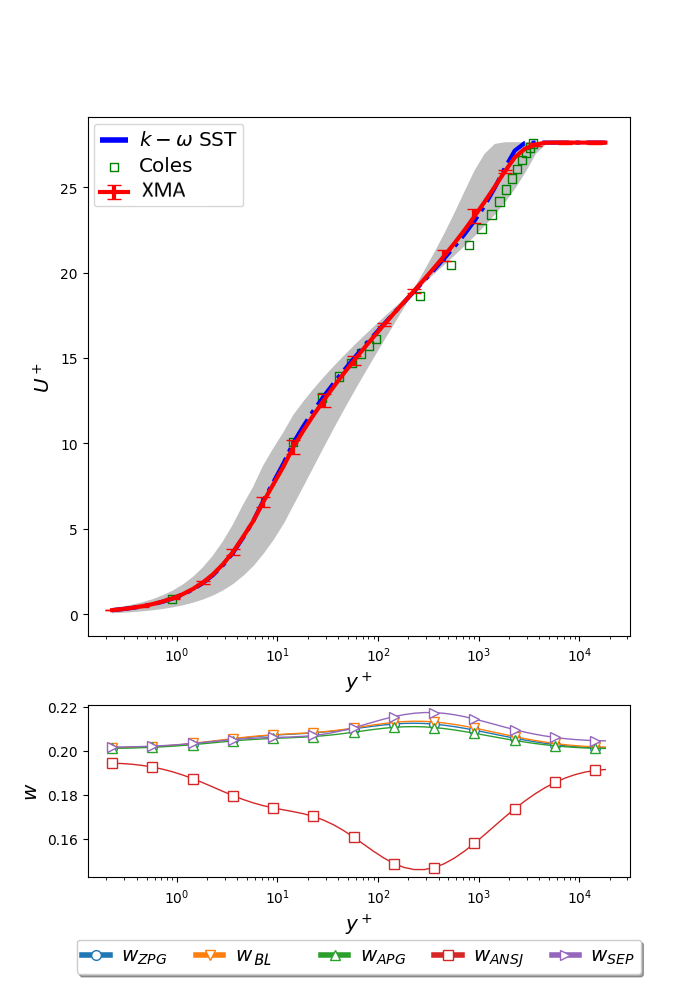}
    \caption{Horizontal velocity (in wall units) $U^+$ versus the non-dimensional normal coordinate $y^+$.}\label{zpgU}
  \end{subfigure}
  \hfill
  \begin{subfigure}[b]{0.49\linewidth}
    \includegraphics[width=1.0\linewidth, trim=0 0 42 0, clip]{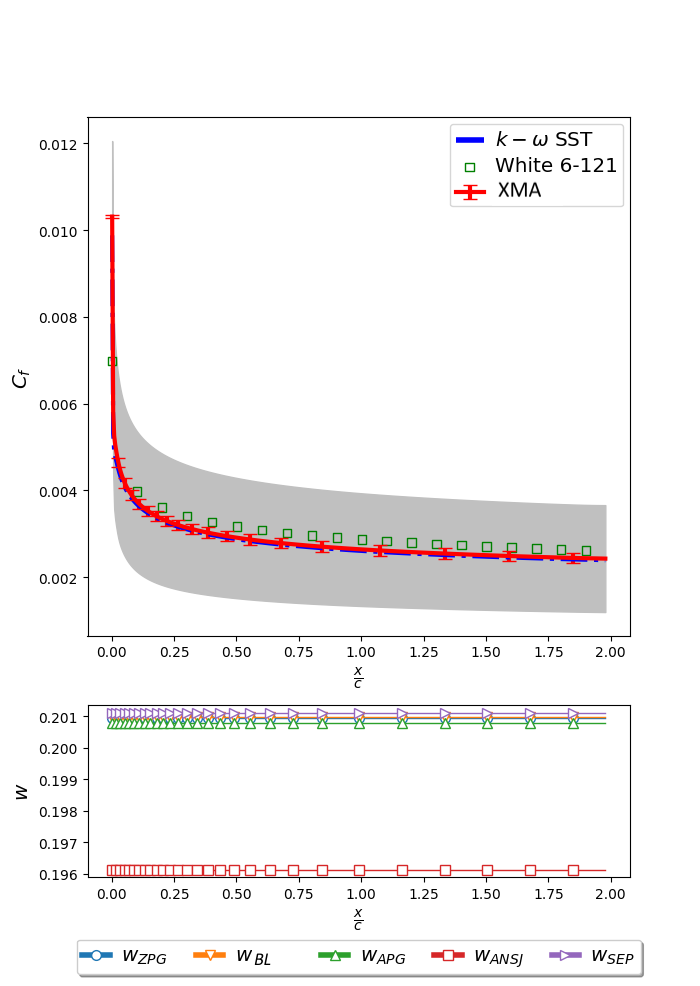}
    \caption{Skin friction $C_f$ distribution along the bottom wall as a function of the non-dimensional streamwise coordinate $x/c$.}\label{zpgCf}
  \end{subfigure}
  \caption{XMA prediction for the NASA turbulent flat plate flow case (2DZP). The bottom panels show the corresponding weighting function distributions. The grey shade represents the accessible region. Error bars correspond to $\pm 3$ standard deviations.}\label{fig:zpg}
\end{figure}

In Figure \ref{zpgU}, we report the velocity profile at the streamwise location $x=0.97$ where $Re_{\theta}\approx 10 000$. The figure shows that the XMA expectancy is in good agreement with both Coles' mean velocity profile \cite{coles1956law,coles1964turbulent} and the baseline $k$-$\omega$ SST prediction. In the figure we also report the accessible region (grey shade) and error bars corresponding to $\pm 3$ standard variations of the XMA prediction. Such intervals are much narrower than the accessible zone because the outlier models (such as the ANSJ model) are assigned low weights, limiting their contribution to the total variance. Models trained on similar cases, e.g., $\M^{(ZPG)}$, $\M^{(CHAN)}$, and $\M^{(APG)}$, exhibit comparable weighting functions (shown in the bottom part). The SEP model, specifically tailored for separated flows, is assigned locally a slightly higher weight in the external part of the boundary layer, as previously observed for the channel flow case. Finally, the ANSJ is assigned low weight in most of the flow. 

Figure \ref{zpgCf} displays the skin friction coefficient distribution along the plate wall and the corresponding weighting functions. The skin friction was neither used for training the weights or the customized models, but it is tightly related to the velocity profiles. It is then interesting to check whether the weighting functions trained on the velocity are still meaningful. Again, the XMA prediction agrees well with the turbulent correlation given in Equation (6-121) of Ref.\,\cite{white2006viscous} and it is close to the baseline calculation. In this case, the models are ranked according to the values that the weighting functions take at the wall location where all models are almost equally weighted, except $\M^{(ANSJ)}$ which is assigned a lower weight. Note that the goal of XMA is not to select a single "best" model in each flow region but to determine an optimal combination of the component models that captures the data. In this case, some models overpredict the skin friction while others underpredict it. The RF captures this behavior and assigns approximately equal weights after excluding the outlier so that, on average, XMA predicts the correct value. Despite a large spread in model predictions, the ANSJ model is assigned again a low enough weight to limit its contribution to the XMA average and variance, which reduces the predictive uncertainty. Overall, the strong agreement with high-fidelity data underscores the robustness of the XMA approach in reproducing canonical flows.

\subsubsection{Axisymmetric Subsonic Jet (ASJ)}
\begin{figure}[H]
  \centering
  \includegraphics[width=0.7\textwidth]{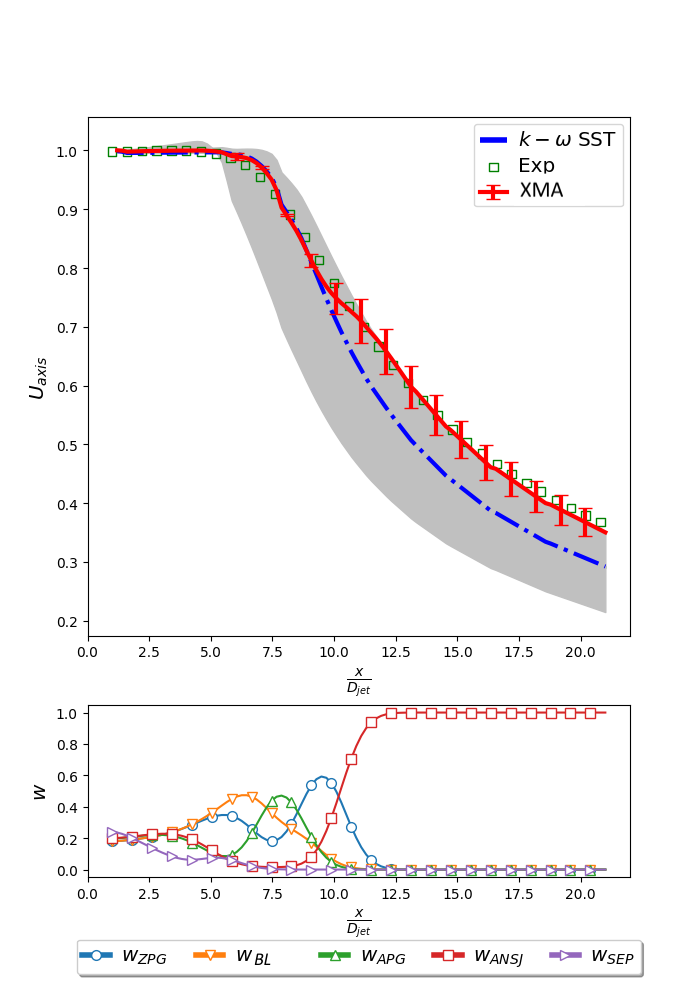}%
  \caption{Distribution of the streamwise velocity $U_{axis}$ along symmetry axis versus the non-dimensional streamwise coordinate $x/D_{jet}$ for the Axisymmetric Subsonic Jet (ASJ) case. The grey shade represents the accessible region. Error bars correspond to $\pm 3$ standard deviations.}
  \label{fig:asj_uaxis}
\end{figure}

\begin{figure}[H]
  \centering
  \centerline{\includegraphics[width=1.2\textwidth, trim= 120 0 120 0,clip]{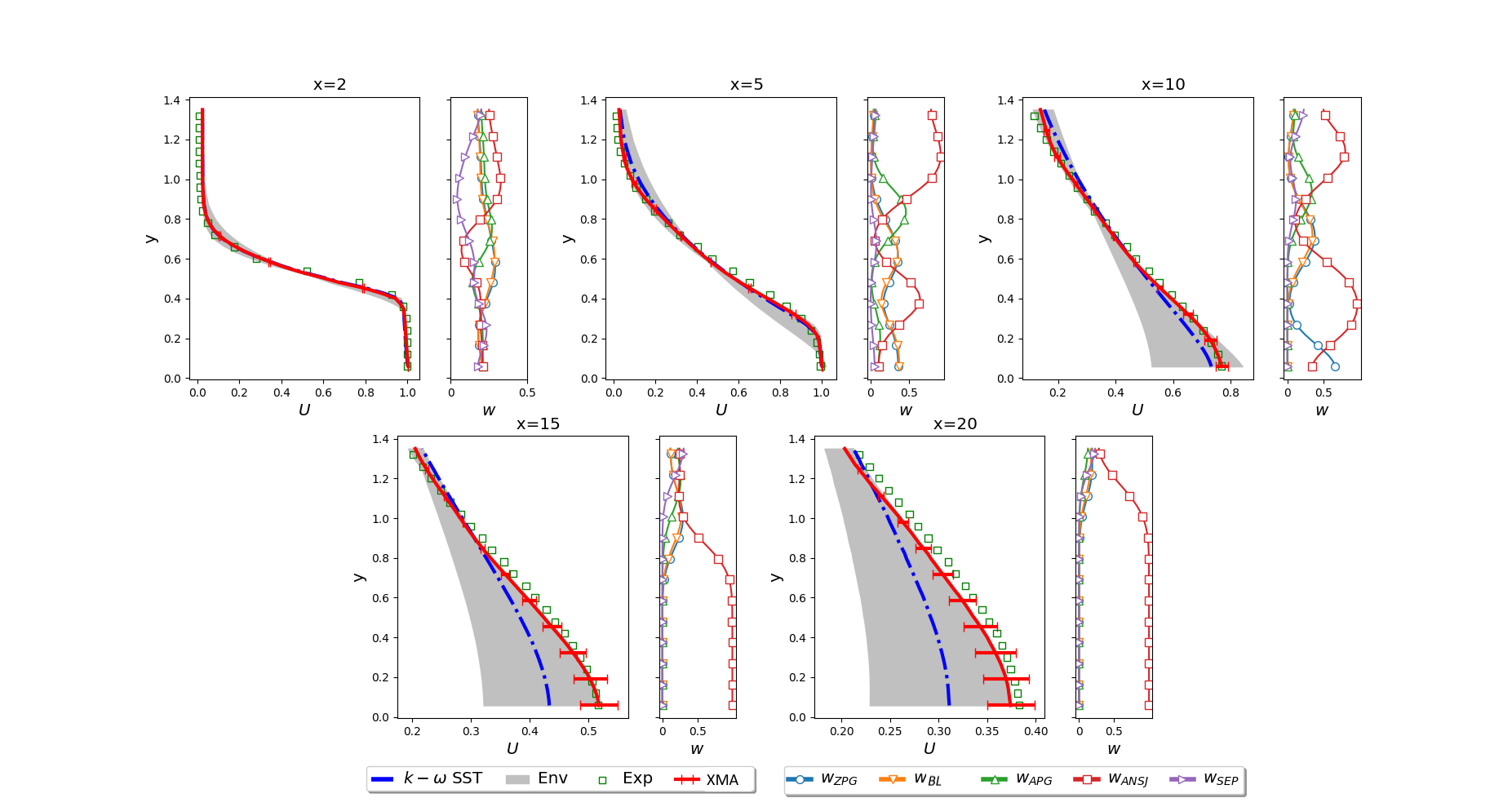}}
  \caption{Profiles of the streamwise velocity $U$ as a function of the crosswise coordinate $y$ at various horizontal locations for the Axisymmetric Subsonic Jet (ASJ) case. The grey shade represents the accessible region. Error bars correspond to $\pm 3$ standard deviations.}
  \label{fig:asj_u}
\end{figure}

Figure \ref{fig:asj_uaxis} displays the distribution of the horizontal velocity $U_{axis}$ along the jet axis. The XMA expected solution shows an excellent agreement with the reference experimental data and represents a significant improvement over the baseline. In the potential region, all component models are assigned approximately equal weights, except for the SEP model which is slightly penalized. In the early jet region there is no clear winner and the mixture smoothly switches from one model to another, until only the ANSJ model emerges in the far jet. The velocity profiles reported in Figure \ref{fig:asj_u} also match very well the reference data, except for the last profile which is not included in the XMA accessible region. In this case, the XMA prediction lies along the boundary closest to the data and its $\pm3$ standard deviations uncertainty bars encompass the reference data. Interestingly, the discrepancies among component models tend to increase when moving downstream. This showcases that, at the farest stations, the ANSJ model performs better than the others yet exhibits a discrepancy with respect to the reference. The large uncertainty bars in the regions of dominance of $w_{ANSJ}$ proves also that this region of the flow is highly sensitive to the parametric uncertainty of this model. 

The ASJ test case is characterized by a notably lower Mach number (by a factor of 5) than the ANSJ case included in the training flow set. Despite this substantial change, the XMA approach continues to yield satisfactory results. Notably, the spreading in the far jet region is accurately captured by XMA, which significantly outperforms the predictions obtained from the baseline $k-\omega$ SST model. The consistency in performance improvements is further confirmed by the data shown in Figure \ref{fig:asj_u}, where the horizontal velocity predicted by XMA remain closely aligned with high-fidelity data. 

\subsubsection{Wall-Mounted Hump (2DWMH)}
\begin{figure}[H]
  \begin{subfigure}[b]{0.49\linewidth}
    \includegraphics[width=1.0\linewidth]{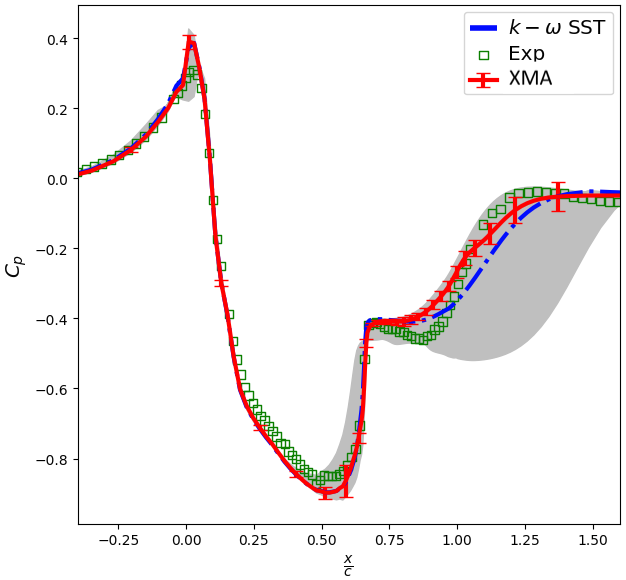}
    \caption{Wall distribution of the pressure coefficient}\label{fig:bumpCp}
  \end{subfigure}
  \hfill
  \begin{subfigure}[b]{0.49\linewidth}
    \includegraphics[width=1.0\linewidth]{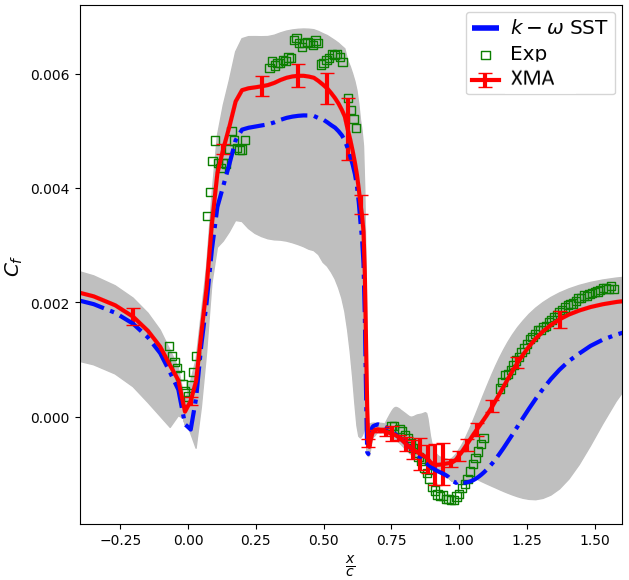}
    \caption{Wall distribution of the skin friction coefficient}\label{fig:bumpCf}
  \end{subfigure} \\
  \begin{subfigure}[b]{1.0\linewidth}
    \centering
    \includegraphics[width=0.59\linewidth]{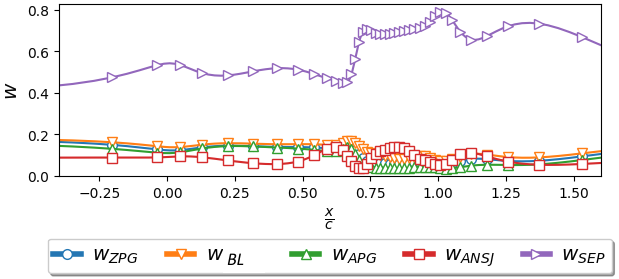}
    \caption{Distribution of the weighting functions versus the non-dimensional streamwise coordinate $x/c$.}\label{fig:bumpw}
  \end{subfigure}
  \caption{NASA 2D Wall-Mounted Hump case (2DWMH): wall distributions of the pressure and skin friction coefficients versus the streamwise coordinate. The grey shade represents the accessible region. Error bars correspond to $\pm 3$ standard deviations. }\label{fig:bumpNASA}
\end{figure}

\begin{figure}[H]
  \centering
  \centerline{\includegraphics[width=1.2\textwidth, trim= 120 0 120 0, clip]{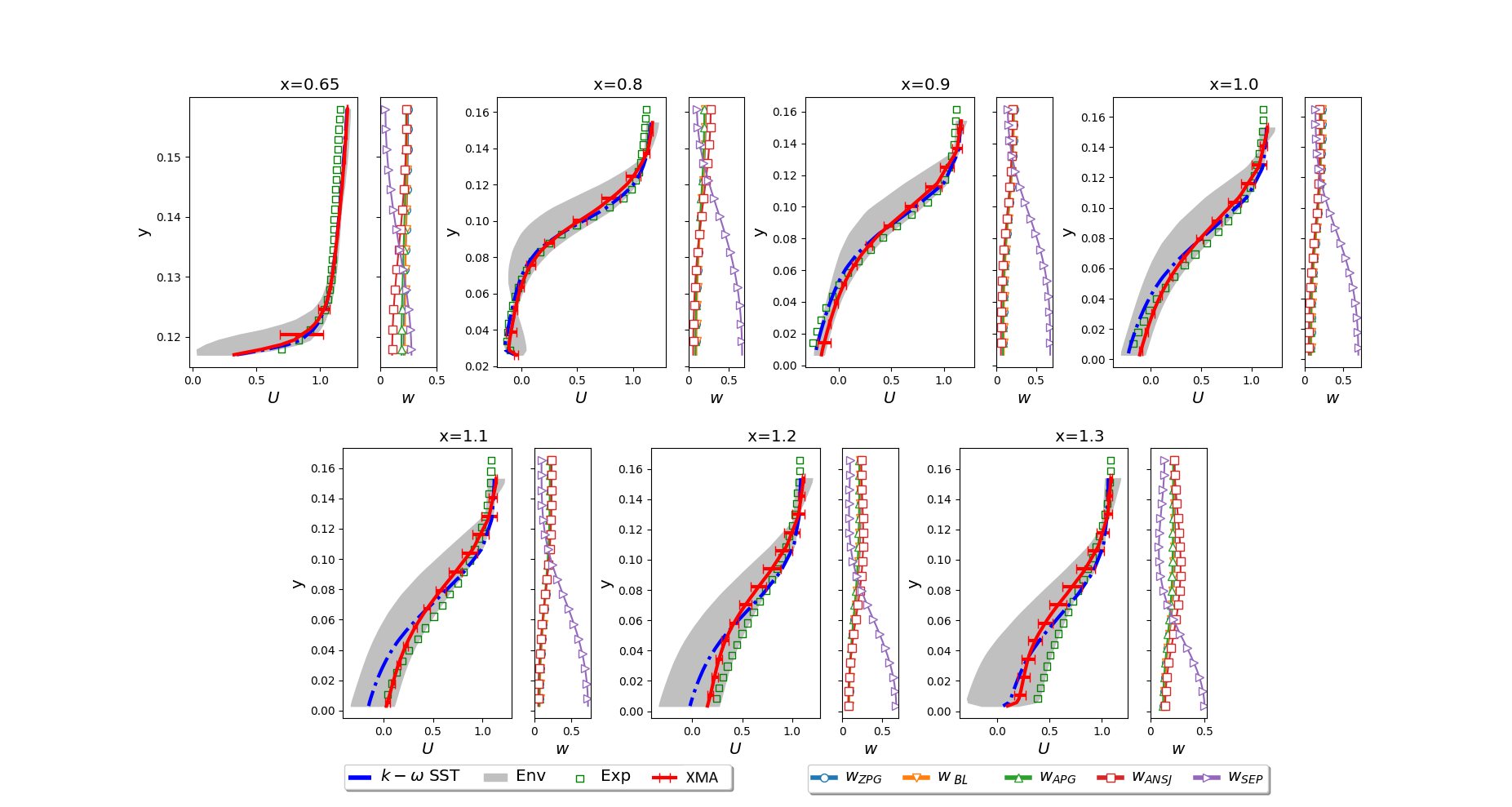}}
  \caption{Profiles of the streamwise velocity $U$ versus the crosswise coordinate $y$ at various streamwise locations for case 2DWMH. The grey shade represents the accessible region. Error bars correspond to $\pm 3$ standard deviations.}\label{fig:bump_u}
\end{figure}

\begin{figure}[H]
  \centering
  \centerline{\includegraphics[width=1.2\textwidth, trim= 120 0 120 0, clip]{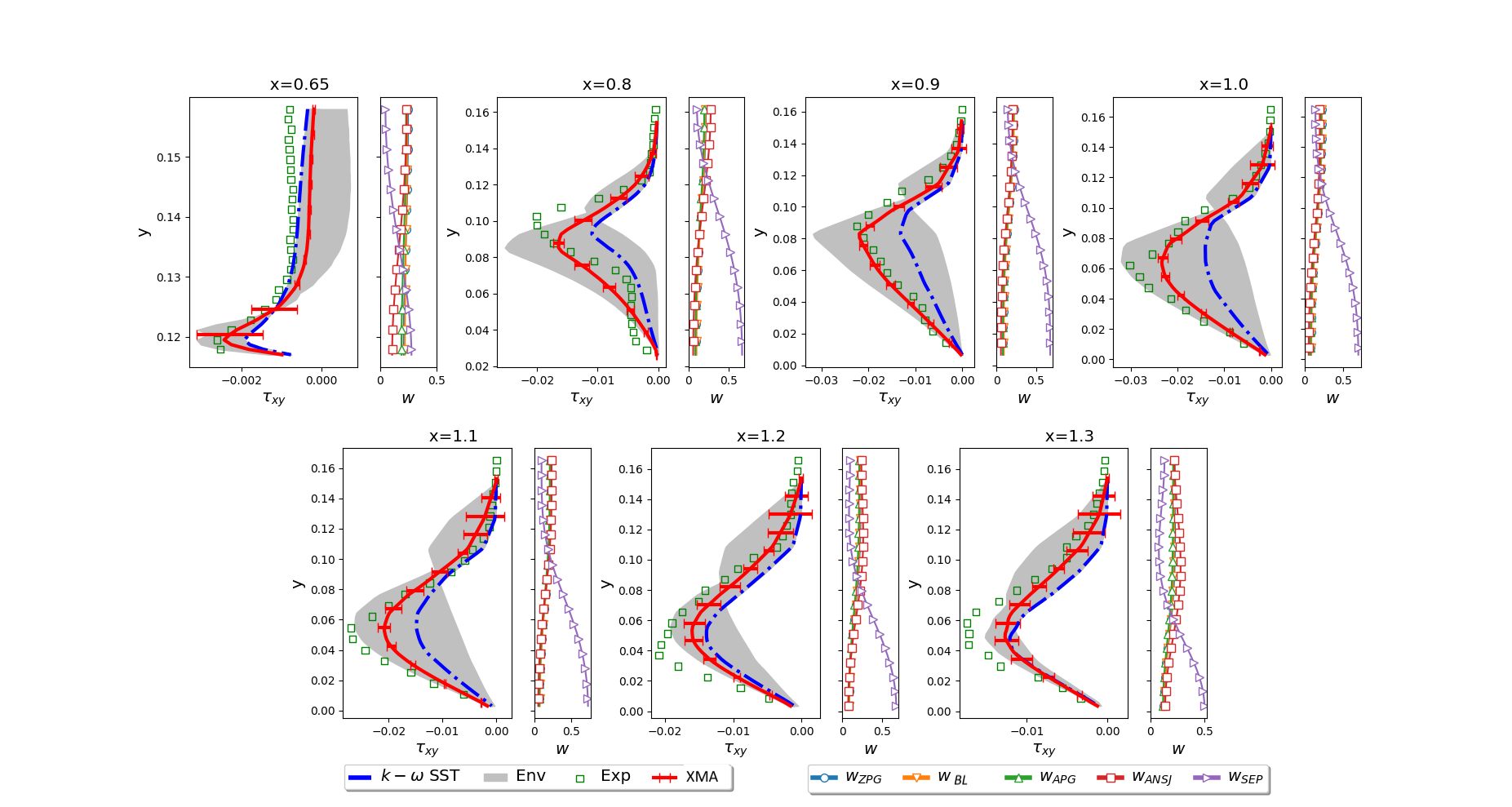}}
  \caption{Profiles of Reynolds shear stress $\tau_{xy}$ versus the crosswise coordinate $y$ at various horizontal locations for the Wall-Mounted Hump case. The grey shade represents the accessible region. Error bars correspond to $\pm 3$ standard deviations.}\label{fig:bump_tauxy}
\end{figure}

The NASA 2D Wall-Mounted Hump is representative of a flow with both attached boundary layer and separation. It operates at a significantly higher Reynolds number ($80 \times 10^6$) than the SEP flow cases used for training (that features Reynolds numbers of the order of $10^4$), and it features a non-equilibrium region downstream of the reattachment point.  Figure \ref{fig:bumpCp} shows the wall distribution of the pressure coefficient $C_p$. The XMA captures very well the high-fidelity data, some discrepancies being visible within the separated region (corresponding to the pressure plateau located approximately between the abscissas 0.75 and 1). The predictions clearly outperform the baseline model in the diverging part. Upstream of the hump, the models designed for flat plates ($\M^{(ZPG)}$ and $\M^{(APG)}$)  are assigned equal weights (reported in Figure \ref{fig:bumpw}), the SEP model emerges as the highest weighted model throughout the flow, whereas the ANSJ model is assigned a low weight. The XMA uncertainty bars are small upstream (except at pressure extrema) and they become larger in the separated region, highlighting the high sensitivity of this region to the injection of eddy-viscosity performed by the SEP model.
In Figure \ref{fig:bumpCf} is reported the wall distribution of the skin friction coefficient $C_f$. XMA captures rather well the reference, showing a great improvement over the baseline and capturing the reattachment location rather well. The weights are the same as for the pressure coefficient, since they only depend on the local flow features and not on the QoI to be predicted. The component models solutions for the $C_f$ are widespread over a large accessible area but, again, the error bars are rather small because only one of the models is assigned a high weight, while the unsuitable model here (the ANSJ model which is responsible for this significant discrepancy in the accessible area) is evidently rated poorly and the three models ($\M^{(CHAN)}$, $\M^{(ZPG}$ and $\M^{[APG)}$) predict similar solutions. Unfortunately, the bars do not always fully encompass the reference solution but they are close to it. Overall, the XMA prediction shows clear improvement compared to the baseline.
In Figures \ref{fig:bump_u} and \ref{fig:bump_tauxy}, we can observe a good agreement between the aggregated XMA solutions for both horizontal velocity $U$ and Reynolds shear stress $\tau_{xy}$ and the high-fidelity data when compared to the baseline model. The agreement is particularly evident in the separated region ($0.8 \leq x \leq 1.2$), where the velocity profiles closely match to the experimental data, in contrast to the baseline model which fails to capture reattachment. With regards to the Reynolds shear stress profiles, the XMA prediction closely approaches the high-fidelity data, surpassing again the performance of the baseline model. 

\subsubsection{Summary of the results and discussion}
To provide an overall picture of the performance of the proposed XMA methodology on unseen flow cases, the improvement metric \eqref {Eq:Imp} is displayed in Table \ref{tab:improvements} for various quantities of interest. Results obtained with the individual expert models are reported alongside the aggregated XMA prediction. The results show that the customized models perform remarkably well for flows similar to those they have been trained on. However, their performance significantly deteriorates for different flows.
Conversely, the XMA prediction consistently outperforms the baseline $k$-$\omega$ SST model for all cases.  More importantly, in all cases but one, the XMA prediction even surpasses the performance of the optimal customized model. This shows that the XMA prediction effectively captures the combined effects of the different customized models used in the mixture by enhancing the prediction locally where the customized model predictions may not be optimal.
\begin{table}[H]
  \centering
  \begin{tabular}{lcccccc}
    \textbf{Case} & \textbf{QoI} & \textbf{XMA} & $\M^{(ZPG)}$ & $\M^{(APG)}$ & $\M^{(ANSJ)}$ & $\M^{(SEP)}$ \\
    \hline\hline
    \multirow{2}{*}{2DZP} & $U$& \textbf{7.0}  & 6.6 & 5.5 & \textit{-927.1} & -241.0 \\
    & $C_f$&\textbf{10.6}  & 6.2 & 4.2 & \textit{-3580.0} & -1868.0 \\ 
    \hline
    \multirow{2}{*}{ASJ} & $U$& \textbf{79.6}  & 8.5 & -13.6 & 72.0 & \textit{-535.9} \\ 
    & $\tau_{xy}$& 13.3  & 5.6 & -21.3& \textbf{51.6} & \textit{-595.1} \\ 
    \hline
    \multirow{4}{*}{2DWMH} & $U$& \textbf{76.0}  & -18.9 & -70.6 & \textit{-1265.8} & 65.4 \\ 
    & $\tau_{xy}$& \textbf{74.1} & -3.8 & 37.0 & \textit{-152.4} & 60.9 \\ 
    & $C_p$& \textbf{17.9} & -6.7 & -36.0 & \textit{-1098.0} & -21.5 \\ 
    & $C_f$& \textbf{64.0} & -18.2 & -9.7 & \textit{-819.2} & 47.6 \\ 
     \hline
  \end{tabular}
   \caption{Improvements in ($\%$) over the baseline $k$-$\omega$ SST for various test cases.}\label{tab:improvements}
\end{table}

\section{Conclusions}
In this study, we presented a machine-learning methodology for learning and aggregating customized turbulence models (experts) for selected classes of flows, in order to make predictions of unseen flows, with quantified uncertainty. 

The expert models are learned using a Bayesian symbolic regression algorithm, and take the form of tangible analytical corrective terms of the  the $k-\omega$ SST model, introduced in the Boussinesq constitutive equation for the Reynolds stress tensor and in the model transport equations. The Bayesian formulation allows to infer posterior probability distributions of the model parameters, which can be propagated through a flow solver using a sparse chaos polynomial method to quantify model uncertainty. 

The expert models learned for various classes of flows (including channel flows, flat plate flows at various Reynolds numbers and pressure gradients, jet flows, and separated flows) perform well for flows within or close to those in the training set but extrapolate badly to very different flows because of opposite correction requirements. A good example is given by the jet flow and the separated flows: for the jet, the correction tends to reduce the model's eddy viscosity while, for the separated flows, terms contributing to increase turbulent dissipation are needed. For flat plate and channel flows (including flat plate flows with adverse pressure gradients), the learned model corrections produce no or little changes with respect to the baseline model.

In order to make predictions of more general flows while estimating the uncertainty associated with the turbulence model choice for extrapolation scenarios, we propose a Mixture-of-Experts approach, named space-dependent model aggregation (XMA). The latter generates a local convex linear combination of a set of expert model solutions by means of weighting functions that depend on a vector of well-chosen flow features. In this study, the weighting functions are expressed as functions the model errors with respect to the reference data: a model is then assigned low weight in regions were it is inaccurate and high weight were it performs well. An input/output relation between a set of local flow features and the model weights is then learned by training a Random Forest model on a set of flow configurations representative of attached flows, separated flows, and free shear. The XMA solution for a new flow is then obtained as a locally weighted average of the solutions of the component models. The model mixture is also used to estimate the predictive variance, \emph{i.e.} a measure of modeling uncertainty. In regions of large discrepancies of the individual solutions, the expert model solutions can spread over a large region, and XMA variance becomes larger. XMA also effectively detects the presence of outlier predictions, which are generally assigned low weights and contribute little to the overall variance. Regions of high variance are then mostly associated to regions where XMA cannot discriminate well enough the component models, thus warning the user about the reliability of the computed prediction. In terms of computational cost, the XMA prediction is more expensive than a standard RANS simulation, because it requires at least one simulation for each expert model, as well as an additional baseline simulation to evaluate the flow features (note that the baseline is also one of the component models). In return,  XMA delivers a measure of model uncertainty. If the probability distributions associated with model parameters are also propagated through the solver, additional simulations are required, the number of which depends on the number of model parameters to be propagated and on the uncertainty quantification algorithm used for the propagation. SBL-SpaRTA models being very sparse (only a few terms are selected by the algorithm), the number of uncertain model parameters to be propagated is small (typically between one and four), and the propagation cost can be greatly reduced by using a sparse uncertainty quantification method.

The XMA approach has been tested not only for flow cases included in the test set, but also to various test cases of the NASA turbulence modeling testing challenge \citep{rumsey2022nasa}, which were not used for training. The results clearly demonstrate that the XMA  activates the most appropriate models in their respective zones of expertise, leading to significantly improved predictions across various QoI. Importantly, the aggregation approach does not compromise the accuracy of predictions for canonical flows where the baseline model already performs well and is then assigned high preference over the expert models.

The proposed approach shows promise for alleviating the overfitting problem observed for most data-driven turbulence models in the literature. It may also potentially provide a criterion for triggering the training of additional/alternative expert models for flow cases exhibiting confidence intervals greater than some acceptable threshold.


\appendix

 \bibliographystyle{elsarticle-num} 
 \bibliography{cas-refs}





\end{document}